\documentclass{JHEP3}

\pdfoutput=1

\usepackage{graphicx}
\usepackage{amssymb}
\usepackage{slashed}
\usepackage{amsmath}
\usepackage{url}

\newcommand{\be}{\begin{equation}}
\newcommand{\ee}{\end{equation}}
\newcommand{\Eq}[1]{Eq.~(\ref{#1})}

\title{Jets with Variable R}

\author{David Krohn \\ Department of Physics, Princeton University, Princeton, NJ 08540  \\ E-mail:  \email{dkrohn@princeton.edu} }

\author{Jesse Thaler \\ Berkeley Center for Theoretical Physics, 
  University of California, Berkeley, CA 94720 \\ Theoretical Physics Group, Lawrence Berkeley 
  National Laboratory, Berkeley, CA 94720 \\ E-mail:  \email{jthaler@jthaler.net}}

\author{Lian-Tao Wang \\ Department of Physics, Princeton University, Princeton, NJ 08540   \\ E-mail:  \email{lianwang@princeton.edu}}

\abstract{We introduce a new class of jet algorithms designed to return conical jets with a variable $\Delta R$ radius.   A specific example, in which $\Delta R$ scales as $1/p_T$, proves particularly useful in capturing the kinematic features of a wide variety of hard scattering processes.  We implement this $\Delta R$ scaling in a sequential recombination algorithm and test it by reconstructing resonance masses and kinematic endpoints.  These test cases show 10 -- 20\% improvements in signal efficiency compared to fixed $\Delta R$ algorithms.  We also comment on cuts useful in reducing continuum jet backgrounds.\footnote{The jet algorithms we describe have been implemented as plugins to the \texttt{FastJet} package~\cite{Cacciari:2005hq,Cacciari:Fastjet}.  These plugins are available from the authors upon request.}}

\begin{document}


\section{Introduction}

The high energy frontier of particle physics is probed by hadronic collisions where hard scattering events often result in colored partons.  Because these partons undergo showering and hadronization, one cannot go directly from detector measurements to the four-momenta of the scattering.  Instead, one must cluster hadrons into jets as an approximation to the short-distance kinematics.  

As jets provide an essential window onto hard scattering processes, much recent work has focused on the procedures used to construct them: jet algorithms.  Many fast, infrared/collinear safe jet algorithms have been developed~\cite{Catani:1991hj,Catani:1993hr,Ellis:1993tq,Dokshitzer:1997in,Wobisch:1998wt,Salam:2007xv,Cacciari:2008gp}.  These algorithms cluster hadrons through different mechanisms (via cones or sequential recombination) and in different orders (hard to soft, soft to hard, or by angle).  To date, however, all jet algorithms used for hadronic collisions return jets of uniform characteristic size $\Delta R = \sqrt{(\Delta \eta)^2 + (\Delta \phi)^2}$ on the $(\eta,\phi)$ plane.  

In this paper, we modify existing jet algorithms to cluster with variable $\Delta R$ cone sizes.  To see how this could be useful, consider a resonance at rest in the lab frame decaying into two partons.  To first approximation, the shower of hadrons resulting from each parton will fall in a circular cone of fixed \emph{angular} size regardless of the orientation of the decay with respect to the beam axis.  However, a jet algorithm which uses a fixed cone size in $(\eta,\phi)$ will not reflect this behavior, as fixed $\Delta R$ corresponds to variable angular size.  As we will see, this can be remedied by letting the cone size of a jet vary as
\be
\Delta R \propto \frac{1}{p_T},
\label{eq:varr}
\ee
where $p_T$ is the transverse momentum of the jet with respect to the beam axis.  

Remarkably, this simple modification to the jet cone size finds applications beyond simple resonance decays, and can be useful in studying more complicated kinematic structures.  The algorithms we present are trivially infrared/collinear safe and are appropriate for use at hadron colliders since \Eq{eq:varr} is invariant under boosts along the beam axis.  Moreover, as we will discuss in Appendix~\ref{app:irsafety}, the fact that jet radii become larger at lower $p_T$ makes these algorithms especially robust against splittings from detector effects and different showering approximations.

This paper is organized as follows.  In Section~\ref{sec:recursive}, we discuss sequential recombination jet algorithms, extensions to variable $\Delta R$, and the particularly useful case specified in \Eq{eq:varr} which we call ``VR''.  In Section~\ref{sec:perform}, we quantify the improvement in signal efficiency by using VR algorithms in three different event topologies:  single resonance decay, multiple resonance decay, and three-body gluino decay.  Section~\ref{sec:rdwb} contains a discussion of background shaping and jet quality cuts.  Section~\ref{sec:conc} contains our conclusions.   Discussions of effective jet radii and VR algorithm parameters can be found in the appendices.  


\section{New Recursive Jet Algorithms}
\label{sec:recursive}
\subsection{Brief Review}
Modern jet algorithms fall into two general categories: cone-based and sequential recombination.  We focus on the latter recursive algorithms because their infrared/collinear safety is easier to prove, although much of  the following discussion could be adapted to cone-based algorithms.

Sequential recombination algorithms begin with a set of four-momenta derived from detector calorimeter cells and then recursively combine pairs of momenta into jets.  To do this, they take a list of initial four-momenta and assign each pair $(i,j)$ a ``jet-jet distance measure'' $d_{ij}$, while each individual four-momenta is assigned a ``jet-beam distance measure'' $d_{iB}$.  At each step, the smallest entry in the set of $d_{ij}$ and $d_{iB}$ is identified.  If the smallest entry is a jet-jet measure, the two four-momenta are combined into one, their prior distance measures removed from the list, and a new entry with the sum four-vector is computed.  If the smallest entry is a jet-beam measure, then the corresponding four-momenta is ``merged with the beam'' and set aside.  The algorithm proceeds in this manner until either all four-momenta are merged with the beam (an inclusive algorithm) or a predetermined distance measure $d_{\rm cut}$ is reached (an exclusive algorithm).  After clustering, an inclusive algorithm returns the four-momenta merged with the beam whose $p_T$ is greater than some minimum value, while an exclusive algorithm returns the unmerged four-momenta (that is, those whose $d_{iB} > d_{\rm cut}$). Here we will focus on inclusive algorithms because they are in more widespread use and because not all algorithms have well-defined exclusive modes.\footnote{For instance, the anti-$k_T$ algorithm \cite{Cacciari:2008gp} assigns smaller $d_{iB}$ to harder jets, so these would be merged with the beam and \emph{not} identified as jets if the algorithm were run with a $d_{\rm cut}$.}

It is possible to parameterize the most popular sequential recombination algorithms for use at hadron colliders via \cite{Cacciari:2008gp}
\be
\label{eq:param}
d_{ij} = \min \left[p_{Ti}^{2n}, p_{Tj}^{2n} \right] R_{ij}^2 , \qquad d_{iB} = p_{Ti}^{2n} \, R_0^2,
\ee
where the values of $n$ for particular algorithms are listed in Table~\ref{tab:corresp}, $R_{ij}$ is the $\Delta R$ separation between the two four-momenta, and $R_0$ is a free parameter that determines the characteristic jet size.  Roughly, $n > 0$ clusters soft items first, $n=0$ clusters by angle, and $n < 0$ clusters from hard particles outward.  We will be particularly interested in algorithms with $n \leq 0$, because as emphasized in Ref.~\cite{Cacciari:2008gp}, such recursive jet algorithms act much like an idealized cone-based algorithm for sufficiently negative values of $n$.

\TABLE[h]{
\parbox{\textwidth}{
\begin{center}
\begin{tabular}{c|c}
\hline 
Algorithm & $n$\\
\hline
\hline
$k_T$ \cite{Catani:1993hr,Ellis:1993tq} & $1$\\
Cambridge-Aachen \cite{Dokshitzer:1997in,Wobisch:1998wt} & $0$\\
Anti-$k_T$ \cite{Cacciari:2008gp} &$-1$ \\
\hline
\end{tabular}
\end{center}
}
\caption{\label{tab:corresp} Parameterization of popular sequential recombination algorithms according to \Eq{eq:param}.}
}

At lepton colliders, \Eq{eq:param} is usually modified by replacing $p_{Ti}$ with the energy $E_{i}$, and $R_{ij}$ with the arc length $S_{ij}$ on the $(\theta,\phi)$ sphere defined by $\Delta S = \sqrt{(\Delta \theta)^2 + (\sin{\theta} \, \Delta \phi)^2}$.   This $\Delta S$ measure will be part of the inspiration for the jet algorithms presented in Section~\ref{sec:vralgs}.

\subsection{Variable $\Delta R$ Algorithms}

We now generalize \Eq{eq:param} so that 
\be
\label{eq:reff}
d_{ij} = \min \left[p_{Ti}^{2n}, p_{Tj}^{2n} \right] R_{ij}^2 , \qquad d_{iB} = p_{Ti}^{2n} R_{\rm eff}(p_{Ti})^2, 
\ee
where $R_{\rm eff}(p_{Ti})$ is a dimensionless number interpreted as an effective jet radius for $n\leq 0$.  Since \Eq{eq:reff} is invariant to boosts along the beam axis, it is appropriate for use at hadron colliders.   See Appendix~\ref{app:irsafety} for more detail on the possible choices for $R_{\rm eff}$ and the restriction to $n \leq 0$.  

To see why $R_{\rm eff}$ is an effective radius for $n \leq 0$, consider the clustering of two four-momenta $i$ and $j$.  These will only be clustered together if 
\be
d_{ij} < d_{iB}, d_{jB}.  
\ee
Let $p_{Ti}> p_{Tj}$.  As discussed in Appendix~\ref{app:irsafety}, in order for $R_{\rm eff}$ to robustly define an effective radius, we must take $d_{iB} < d_{jB}$.  Therefore, the four-momenta are clustered if
\be
\frac{d_{ij}}{d_{iB}} =  \frac{R_{ij}^2}{R_{\rm eff}(p_{Ti})^2} < 1,
\ee
which requires the $\Delta R$ distance between $i$ and $j$ to be within an effective radius $R_{\rm eff}$.

To our knowledge, all current algorithms at hadron colliders use a constant $R_{\rm eff}$.  Although we will only explore one new algorithm here, in which $R_{\rm eff}\propto 1/p_T$, it is possible to invent new algorithms tailor-made to distinct processes.  We leave such extensions to future work, though some guidelines for choosing $R_{\rm eff}$ appear in Appendix~\ref{app:irsafety}.   In principle, a momentum-dependent effective jet size could be implemented in a cone-based algorithm, though we expect that the process of finding stable cones (as is required in SIS-Cone~\cite{Salam:2007xv}) would be much more complicated.  

\subsection{Introducing VR Algorithms}
\label{sec:vralgs}

We now introduce a particular example of the variable $\Delta R$ algorithms described above which we will denote as ``VR''.  To motivate this setup, consider a resonance decaying in its center-of-mass frame.   If the resonance decays to two jets, these jets will be naturally described by circles on the $(\theta,\phi)$ sphere.   This is the setup used to describe jets at $e^+/e^-$ colliders where the center-of-mass frame  \emph{is} the lab frame.  Unfortunately, the partonic center-of-mass frame at a hadron collider is not fixed, and to maintain boost invariance one must work with circles in $(\eta,\phi)$.  This setup can present a ``Goldilocks'' problem in choosing jet radii.   If one is forced to use a fixed $\Delta R$ radius, then the cones in the central region will be large in $\Delta S$, while those in the forward region will be small.  Getting the radius just right will necessarily involve a tradeoff.  

Remarkably, the simplest application of the general jet algorithms described above can remedy this situation, allowing one to use boost invariant jets with a cone size reflecting their true size as measured in the resonance rest frame.  This new VR algorithm works by letting the effective radius of a jet go as
\be
\label{eq:vrscaling}
\mbox{``VR''}:  \quad R_{\rm eff}(p_T) =\frac{\rho}{p_T},
\ee
where $\rho$ is a dimensionful constant.  

To see why this captures the desired behavior, again consider the decay of a resonance into two partons which shower into jets.  In the rest frame of the mother particle, the energy $E$ of each jet is fixed by the mass of the resonance with $E = m_{\rm res.}/2$.  The resulting jets should have roughly the same characteristic angular size $\Delta S$, therefore, the quantity $E \Delta {S}$ is approximately the same for both jets.  As we will now show, fixed $E \Delta {S}$ is equivalent to \Eq{eq:vrscaling} to first approximation.

First note that $E \Delta {S}$ is a quantity that is approximately invariant under both transverse and longitudinal boosts.  Intuitively, for small angles, the invariant mass of two massless four-momenta is equal to the geometric mean of their energies multiplied by their angular separation.  Since invariant mass is a boost invariant quantity, so is $E \Delta S$.  More formally, for small angles
\be
\label{eq:ESequalsPTR}
E \Delta S \approx p_T \Delta R,
\ee
where we have used the fact that $E \approx p_T \cosh \eta$ (true for small jet mass) and $\Delta S \approx \Delta R/ \cosh\eta$ (see Appendix~\ref{app:applic}).  Since $p_T$ and $\Delta R$ are invariant under boosts along the beam axis, so is $E \Delta S$ to first approximation.   We emphasize that this is true even for boosts transverse to the beam axis; just define a new $(\eta,\phi)$ coordinate system along the boost axis and go through the same procedure to show that $E\Delta S$ is invariant.

Putting this all together, we want build a jet algorithm that captures the fact that the two jets have constant opening angle in their mother's rest frame.  Since $E \Delta S$ is approximately boost invariant, this is equivalent to forming jets of constant $E \Delta S$ in any convenient frame.  In particular, in the lab frame we can use \Eq{eq:ESequalsPTR} to swap $E \Delta S$ for $p_T \Delta R$.  Therefore, to get jets of constant $E \Delta S$, we should choose $R_{\rm eff}$ to scale like $1/p_T$ as in \Eq{eq:vrscaling}.   This defines the VR jet algorithms.

From this logic, we expect the parameter $\rho$ to be proportional to the typical jet size measure $E\Delta S$, and thus proportional to the resonance mass $m_{\rm res.}$.   For a more detailed discussion of the valid parameter range for $\rho$, see Appendix~\ref{app:applic}.

\subsection{Event Topologies with VR-symmetry}
\label{sec:vrsymmetry}

The VR scaling of \Eq{eq:vrscaling} is applicable whenever there is reason to expect all jets in an event to have the same $E \Delta S$ in some frame, and we call these events ``VR-symmetric''.  This is certainly the case for a single resonance decay.  Less obvious is that this is true for longer cascade decays; even if a cascade involves many intermediate states, it will still be VR-symmetric as long as all jets come from the decay of resonances with a common mass.   VR-symmetry can even be satisfied when there is no actual reconstructable resonance.  For example, the three-body gluino decay $\tilde{g}\rightarrow 2j+\tilde{\chi}_1^0$ would satisfy the requirement toward the kinematic endpoint.  We will discuss these scenarios in more detail in the next section.

An important example without VR-symmetry is initial state radiation (ISR).  Jets from ISR do not have a preferred mother rest frame and so the VR jet cone scaling is not appropriate.  In the case of resonance production plus ISR, the hardest two jets will have the VR scaling, but the ISR will not, so in principle, a hybrid VR/fixed-cone algorithm could have better jet reconstruction performance.  While we will include ISR in our Monte Carlo simulations, we will only study the hardest jets in an event for which VR-symmetry is expected to apply.

Now we would like to address a few caveats to the VR derivation in Section \ref{sec:vralgs}. We derived our expression for $R_{\rm eff}$ in the small cone limit.  In practice, one must account for corrections in going to finite-sized jets when choosing reasonable jet parameters (see Appendix~\ref{app:applic}).   Similarly, for low $p_T$ the algorithm would return pathologically large jets, so one is forced to cut off the jet radius at some $R_{\rm max}$, so that 
\be
\label{eq:finalReff}
R_{\rm eff}(p_T)=\min\left[\frac{\rho}{p_T},R_{\max}\right].
\ee
For small enough $R_{\max}$, \Eq{eq:finalReff} effectively defines a kind of hybrid VR/fixed-cone algorithm as needed for events involving ISR, though in this paper we will not try to optimize the value of $R_{\max}$.

As a secondary issue in the VR derivation, we implicitly assumed that jets will remain circular in $\Delta R$ as we boost from the mother rest frame to the lab frame.  While this is true if the boost is only along the beam axis (as is the case for single resonance production), for more complicated event topologies, the particular shape of a jet will depend on the orientation of the hard process, and will in general be non-circular.  Our algorithm forms circular jets in $(\eta,\phi)$ so it will not capture the true jet shape in more complicated decays.  However, unlike a fixed $\Delta R$ algorithm, the VR algorithms do scale the overall jet size appropriately under boosts.  Since jets are only conical in a statistical sense anyway, we do not expect shape distortions to reduce the efficacy of the VR algorithms.

We can implement \Eq{eq:finalReff} in any existing jet algorithm parameterized by \Eq{eq:reff}. However, as discussed in Appendix~\ref{app:irsafety}, it is only meaningful to define $R_{\rm eff}$ when $n\leq 0$.  Therefore, it does not make sense to apply it to the $k_T$ algorithm where $n=1$.  We will therefore combine it with anti-$k_T$~\cite{Cacciari:2008gp} (denoted AKT) and Cambridge-Aachen~\cite{Dokshitzer:1997in,Wobisch:1998wt} (denoted CA), and compare the resulting algorithms, AKT-VR and CA-VR, with the original AKT and CA algorithms.  A summary of the jet definitions used  in our study appear in Table~\ref{tab:jetsummary}, and a sample lego plot showing the effects of the VR algorithm in reconstructing an event can be seen in Figure~\ref{fig:lego}.

\TABLE[h]{
\parbox{\textwidth}{
\begin{center}
\begin{tabular}{c|c|c}
\hline 
Shorthand & $n$ & $R_{\rm eff}$ \\
\hline
\hline
AKT &$-1$ & $R_0$\\
AKT-VR &$-1$ & $\rho/p_T$\\
\hline 
CA & $0$ & $R_0$ \\
CA-VR & $0$ & $\rho/p_T$\\
\hline
\end{tabular}
\end{center}
}
\caption{\label{tab:jetsummary} The four jet algorithms used in this study, as parameterized by \Eq{eq:reff}.  For the VR algorithms, we also impose a maximum jet radius as in \Eq{eq:finalReff}.}
}

\FIGURE[t]{
\includegraphics[scale=0.4]{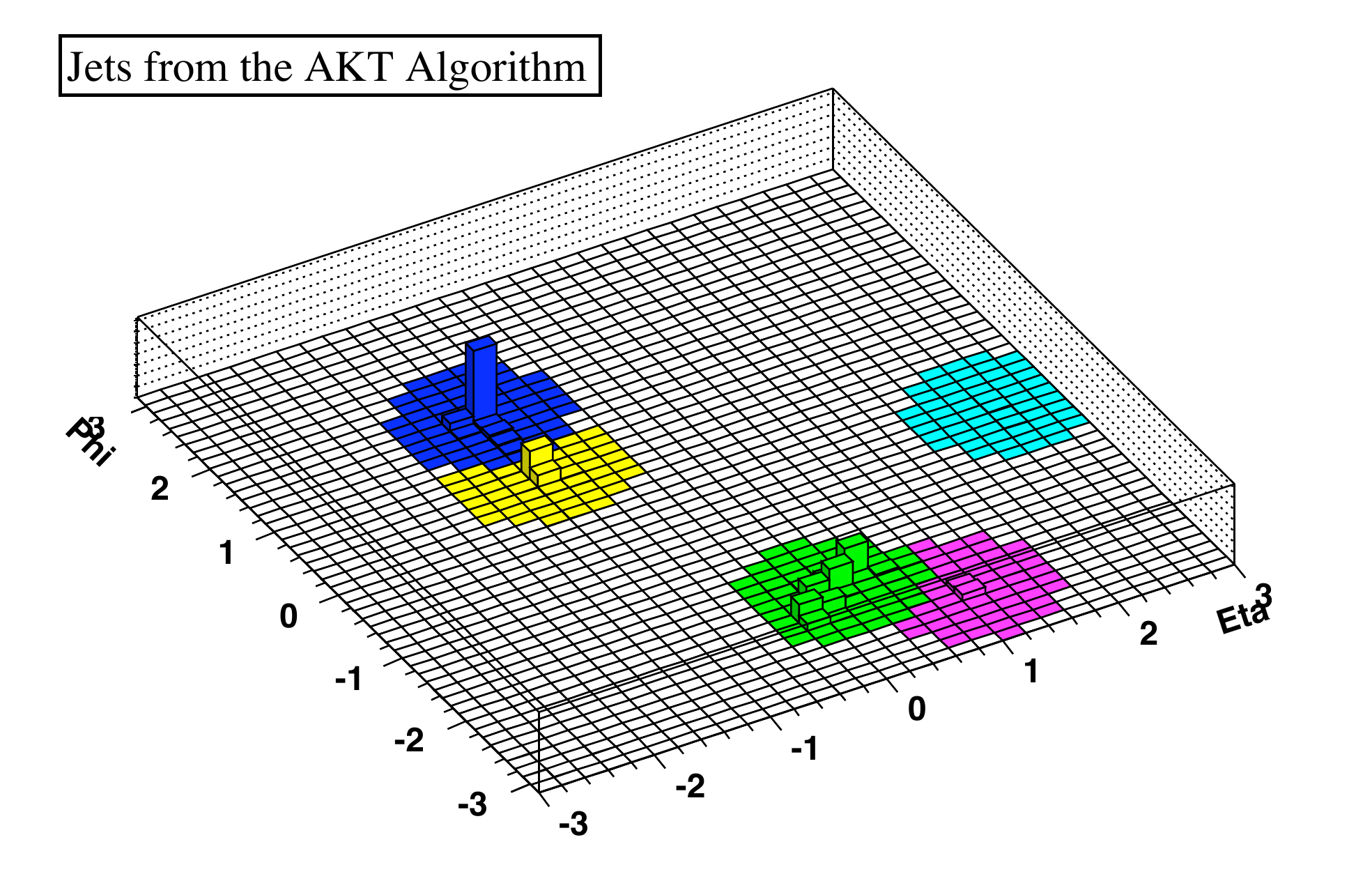}
\includegraphics[scale=0.4]{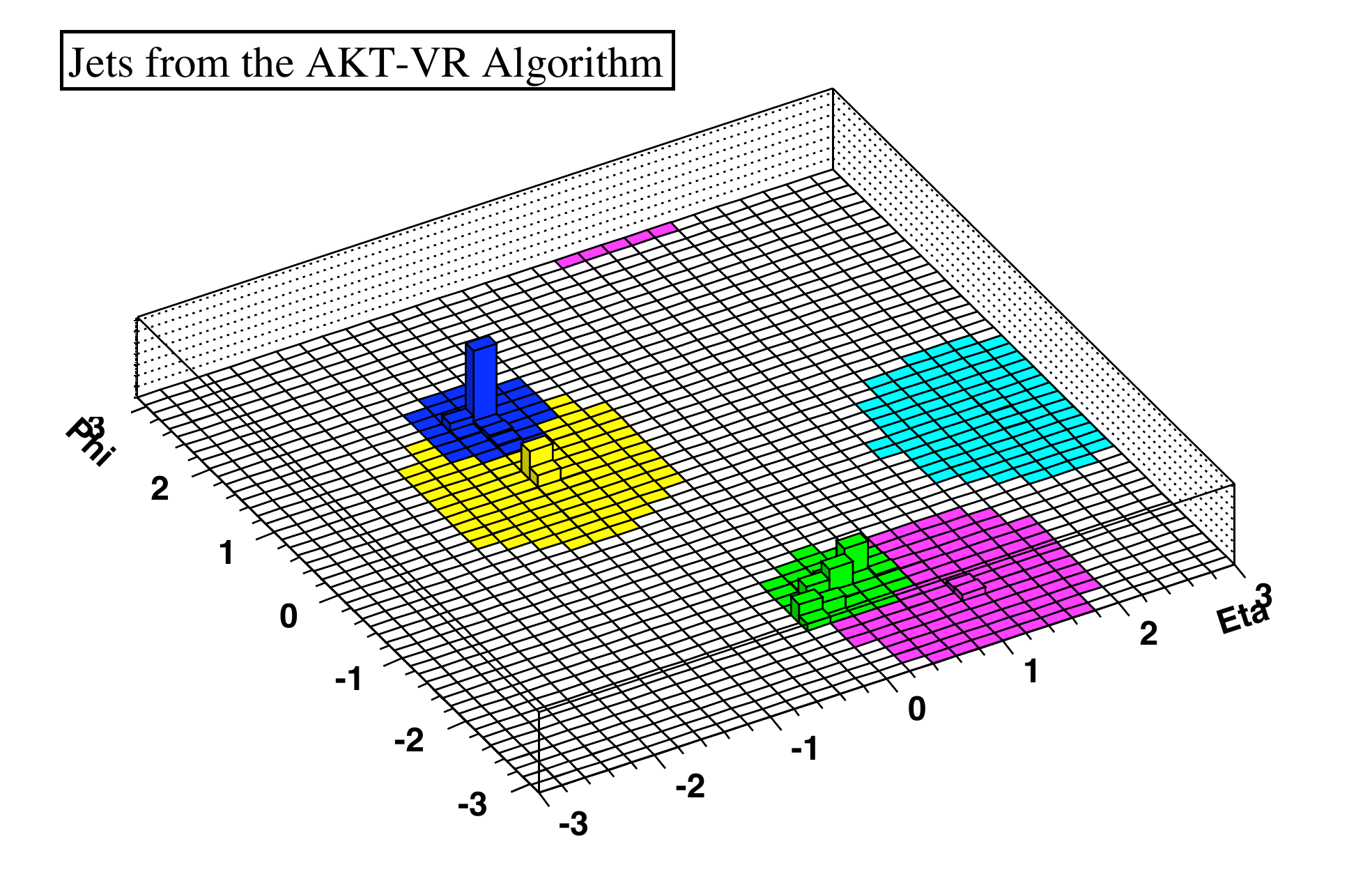}
\caption{\label{fig:lego}The same event reconstructed by anti-$k_T$ (left) and its VR modification (right).  Note that in going to the VR algorithm, the high-$p_T$ jets (dark blue, green) have been reduced in size while softer jets (yellow, purple, light blue) have grown.  In this example, only the two harder jets are expected to exhibit VR-symmetry, and the softer jets are saturating the $R_{\rm max} = 1.0$ constraint.}
}


\section{Jet Reconstruction Performance}
\label{sec:perform}

Here we use Monte Carlo simulations to compare the AKT and CA algorithms with their VR cousins.  We investigate three different kinematic scenarios to see how the algorithms perform with both large and small jets.  We focus only on signal efficiency in this section, and turn to background rejection issues in Section~\ref{sec:rdwb}.  

Our signal and background samples have both been generated in \texttt{Pythia 6.4.14}~\cite{Sjostrand:2006za}, with parton-level signal events generated in \texttt{MadGraph 4.4.5}~\cite{Maltoni:2002qb}.  We use nominal LHC beam parameters (14 TeV proton-proton collisions).  Final state hadrons are grouped into $\delta \eta \times \delta \phi = 0.1\times 0.1$ calorimeter cells between $-3 < \eta < 3$ and assigned massless four-momenta based on the calorimeter energy.  These calorimeter cells are the starting point for the recursive jet clustering.  

We use the \texttt{FastJet 2.3.4} \cite{Cacciari:2005hq,Cacciari:Fastjet} package for the AKT and CA algorithms, and we wrote new \texttt{FastJet} plugins for the AKT-VR and CA-VR algorithms.   For each kinematic scenario, we scan over a range of jet parameters to optimize the jet algorithm performance.  To keep the comparison fair, we limit the maximum effective $\Delta R$ of the VR jet cones using $R_{\max}$ as in \Eq{eq:finalReff}, and scan the $R_0$ parameter of the fixed cone algorithms from $0$ to $R_{\max}$.  In the three cases below, we find a universal improvement in using the VR algorithms over their fixed $\Delta R$ cousins.

\subsection{Resonance Decays Without Background}
\label{subsec:resonancenobackground}

The simplest test of a jet algorithm is resonance reconstruction without standard model background.  We consider resonances with backgrounds in Section~\ref{sec:rdwb}.  Here we consider the scenario of a color-octet scalar $X$, of negligible width, in the process $gg\rightarrow X\rightarrow gg$.\footnote{The $X$ couples to gluons via the operator $\mathrm{Tr}(X G_{\mu\nu}G^{\mu\nu})$.}  We scan the jet parameters up to a maximum radius $R_{\max} = 1.5$, and optimize the parameters to maximize the percentage of events reconstructed in a narrow mass window ($m_X \pm 25~\mathrm{GeV}$) around the true resonance mass.\footnote{The $\pm25~{\rm GeV}$ mass window was chosen by hand to approximate the width of the reconstructed peaks after showering and hadronization.  It is not related to the perturbative resonance width, which is zero, or calorimeter smearing, whose effects we have not included.} The results of this optimization are shown in Table~\ref{tableresnobg} for four different values of $m_X$.

\TABLE[h]{
\begin{tabular}{c|cccc}
\hline 
Algorithm & $500~\mathrm{GeV}$ & $1~\mathrm{TeV}$ & $2~\mathrm{TeV}$ & $3~\mathrm{TeV}$\\
\hline
\hline
AKT $\rightarrow$ AKT-VR      &$18\%$ $(0.9,200)$& $14\%$ $(1.0,450)$ & $10\%$ $(1.2,1000)$& $8\% $ $(1.3,1500)$\\
CA   $\rightarrow$ CA-VR         & $17\%$ $(0.9,175)$ & $14\%$ $(1.0,400)$& $7\%$ $(1.2,1000)$& $9\% $ $(1.2,1500)$\\
\hline
\end{tabular}
\caption{Percentage increase in the number of events reconstructed in the mass window $m_X \pm 25~\mathrm{GeV}$ for the VR variant over the original algorithm.  The numbers in parenthesis are the optimized parameters for the original and VR variant ($R_0$ and $\rho$, with $\rho$ in GeV) respectively. We see that the effective $\Delta R \simeq \rho/m_{X}$ of the VR algorithms is comparable to the fixed $\Delta R$.}
\label{tableresnobg}
}

\FIGURE[tp]{
\includegraphics[scale=0.35]{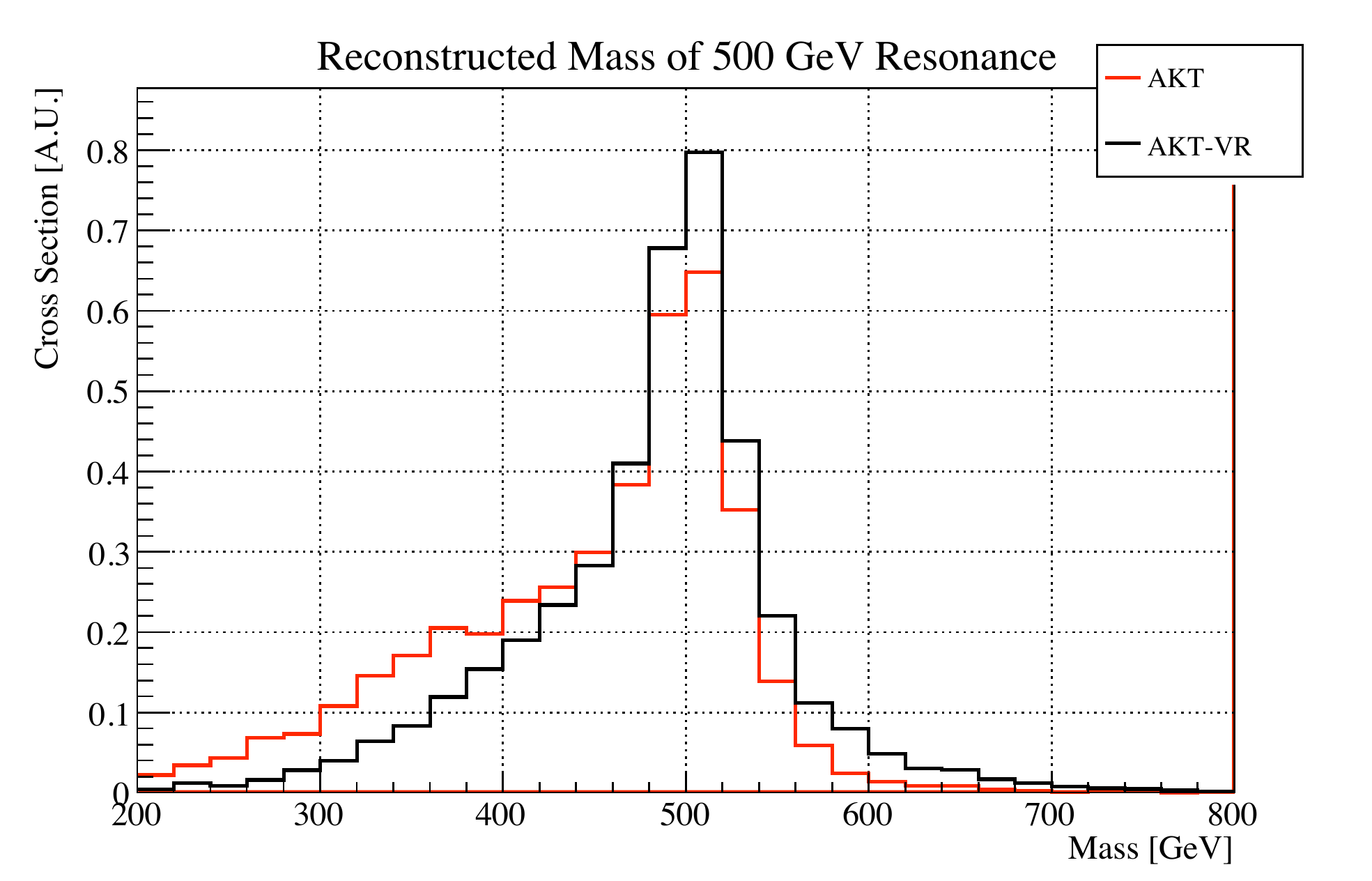}
\includegraphics[scale=0.35]{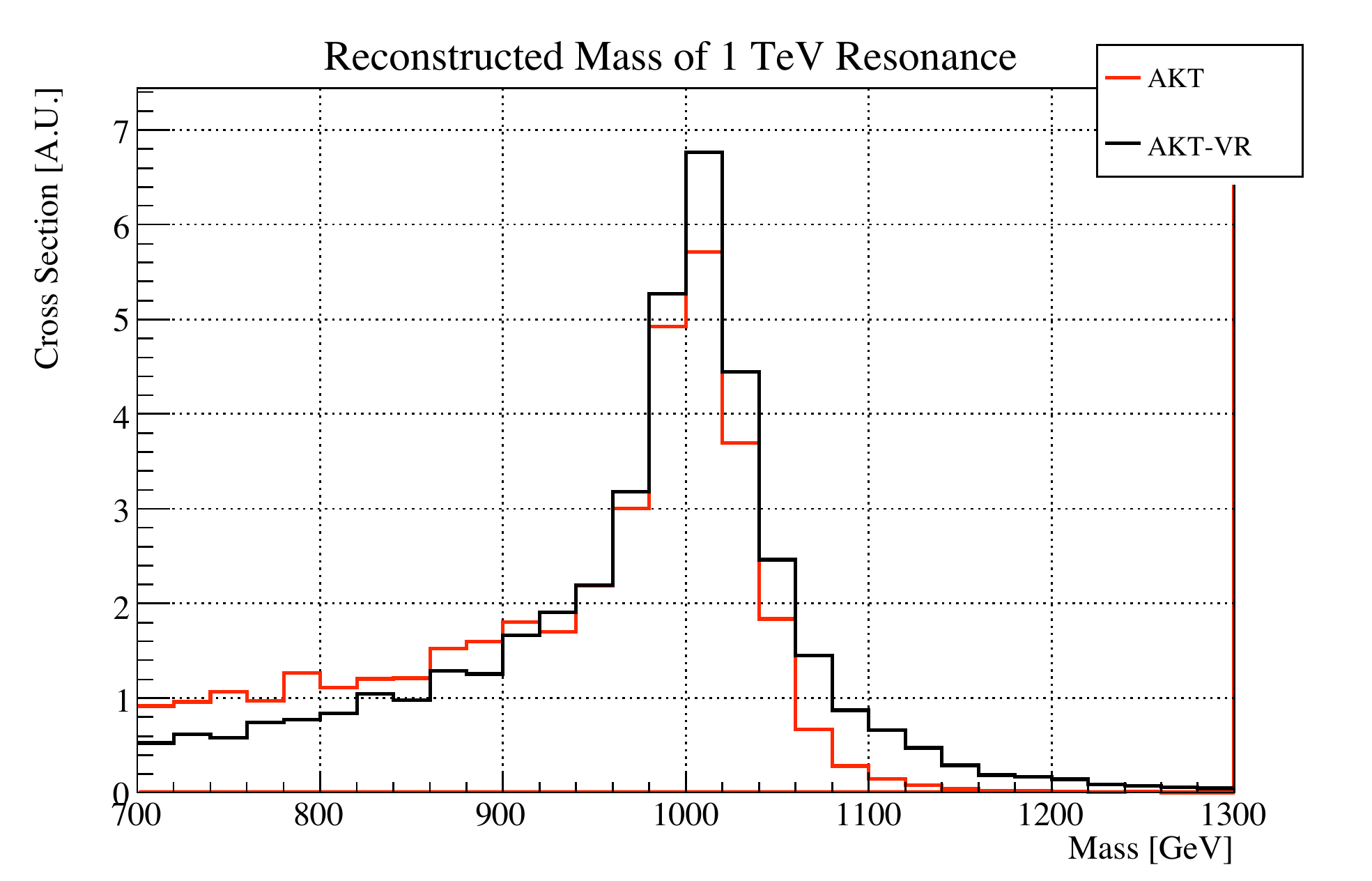}
\\
\includegraphics[scale=0.35]{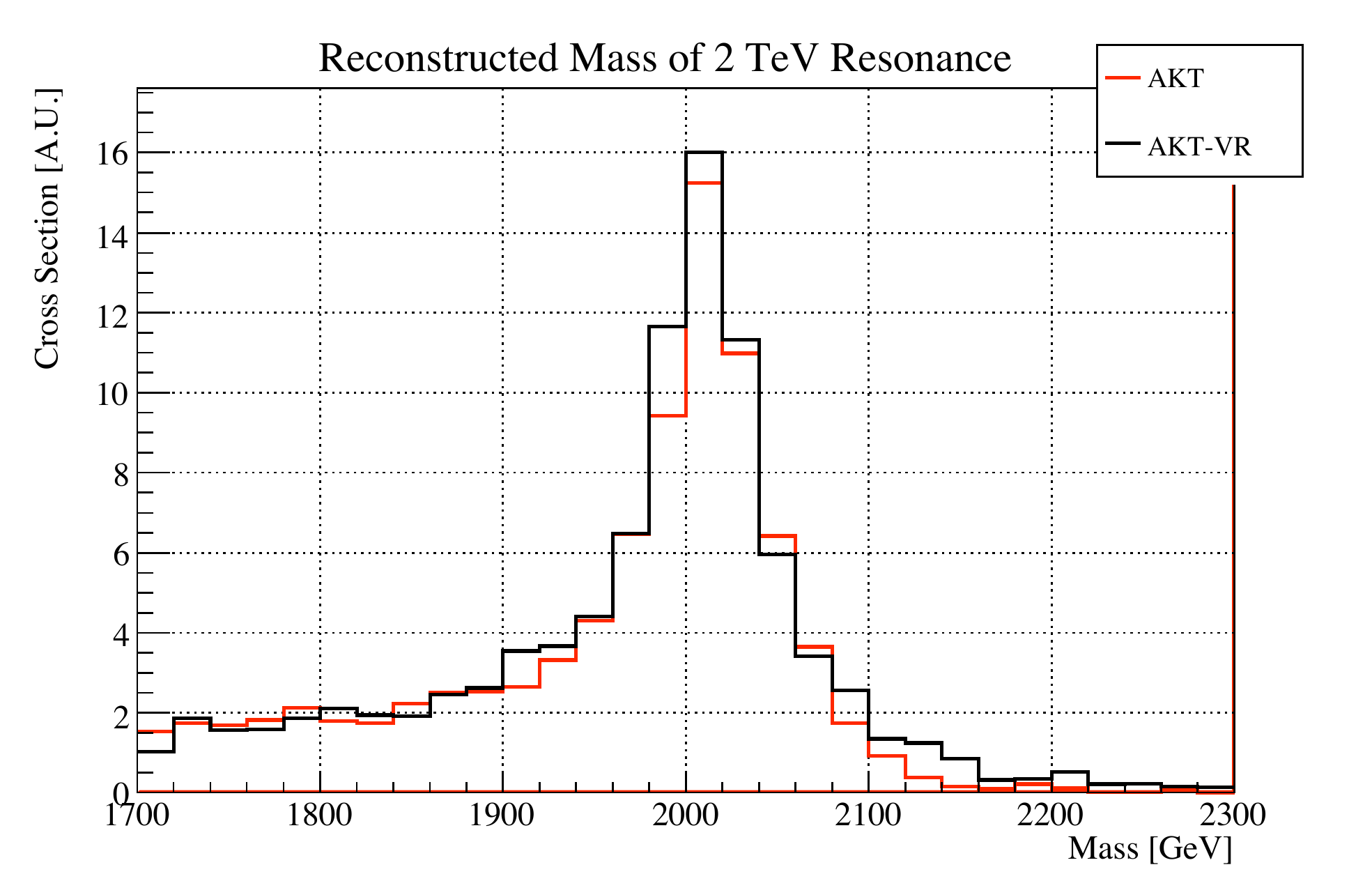}
\includegraphics[scale=0.35]{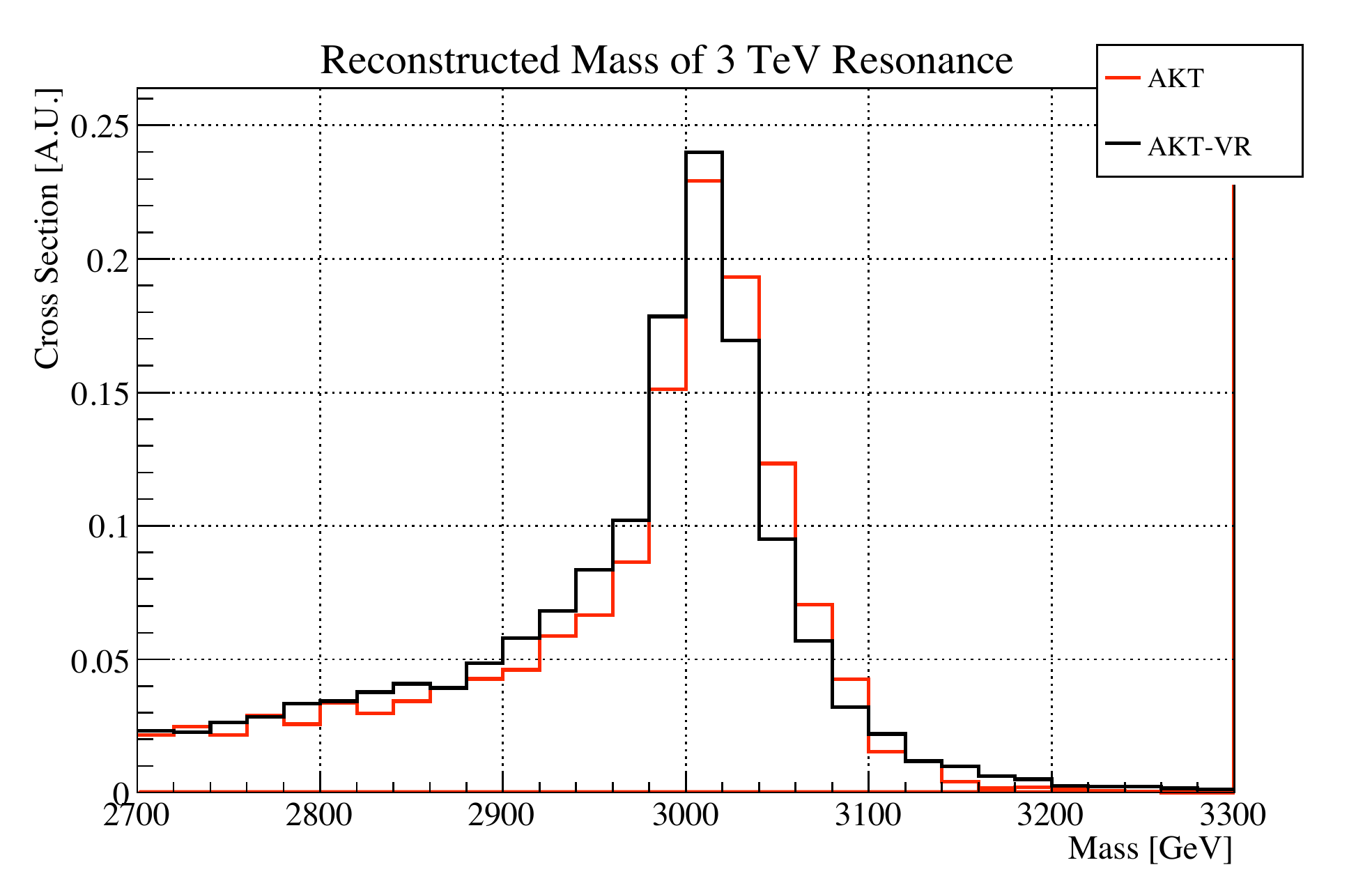}

\caption{\label{fig:resnobgAKT}Invariant mass distributions for the $X \rightarrow gg$ resonance decay using the optimized AKT and AKT-VR parameters in Table~\ref{tableresnobg}.  The distributions have the same normalization, and the $y$-axis is in arbitrary units (A.U.). The VR algorithms yield a better reconstruction, both in the height and width of the resonance.}
}

\FIGURE[tp]{
\includegraphics[scale=0.35]{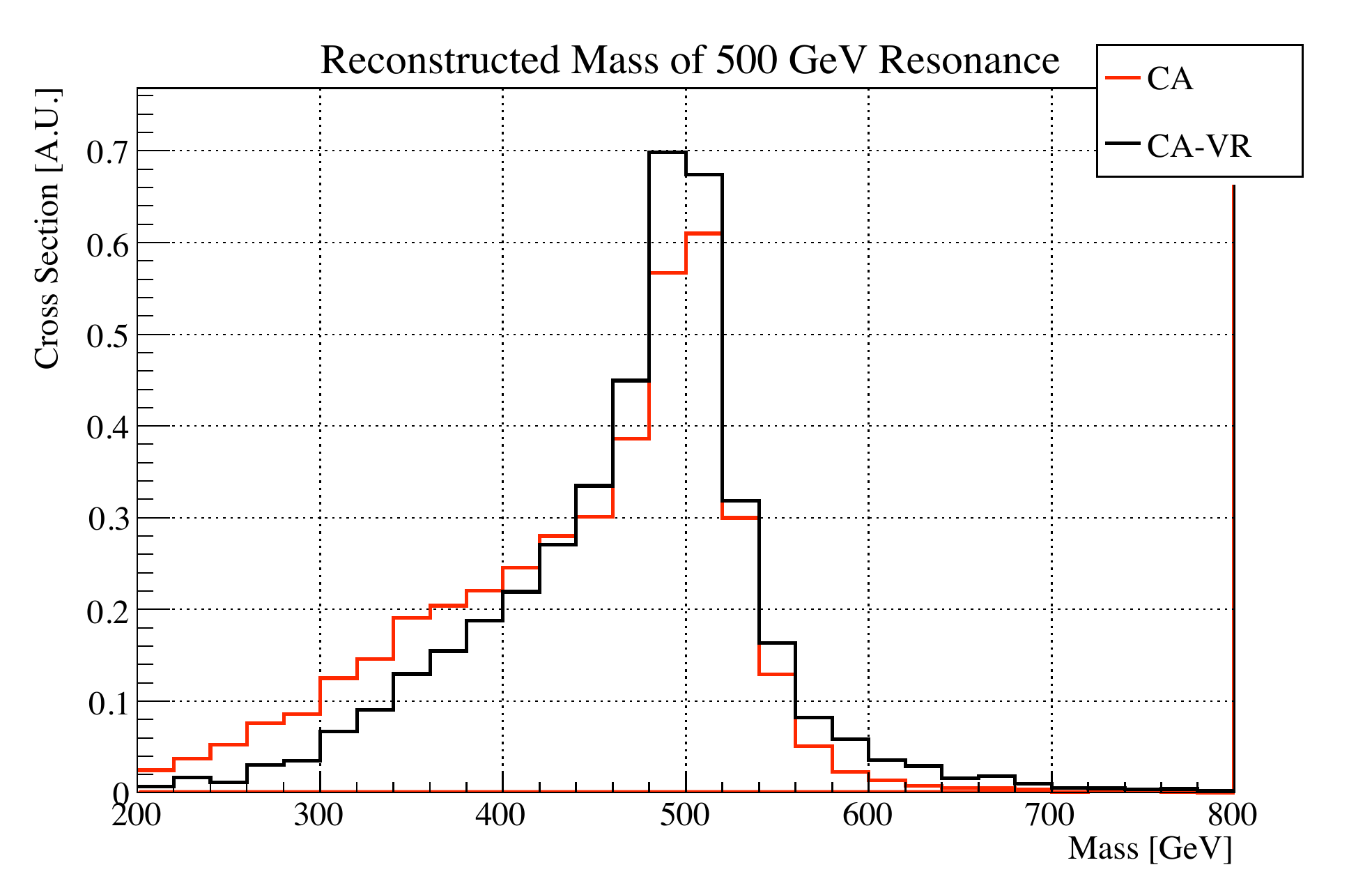}
\includegraphics[scale=0.35]{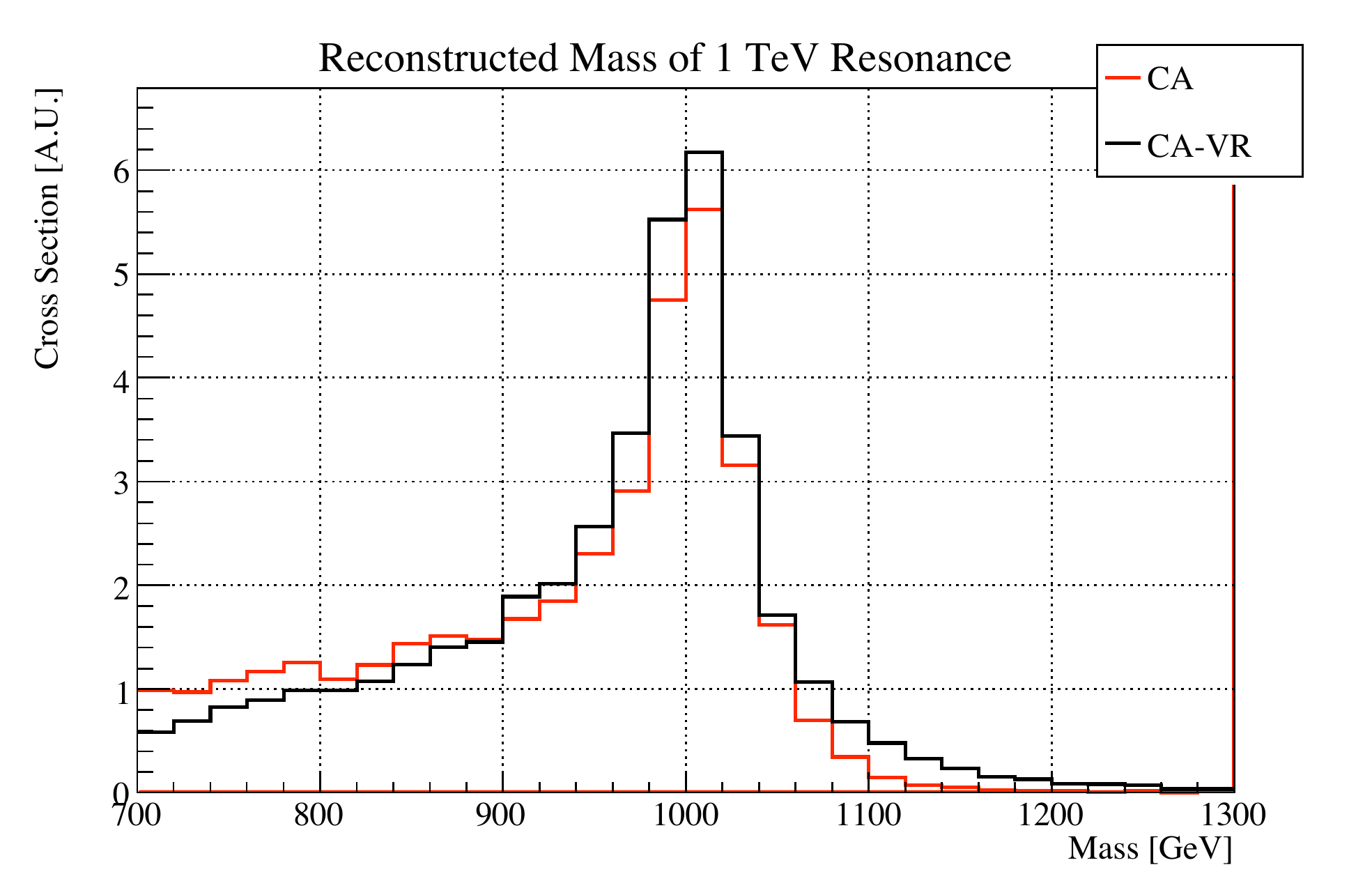}
\\
\includegraphics[scale=0.35]{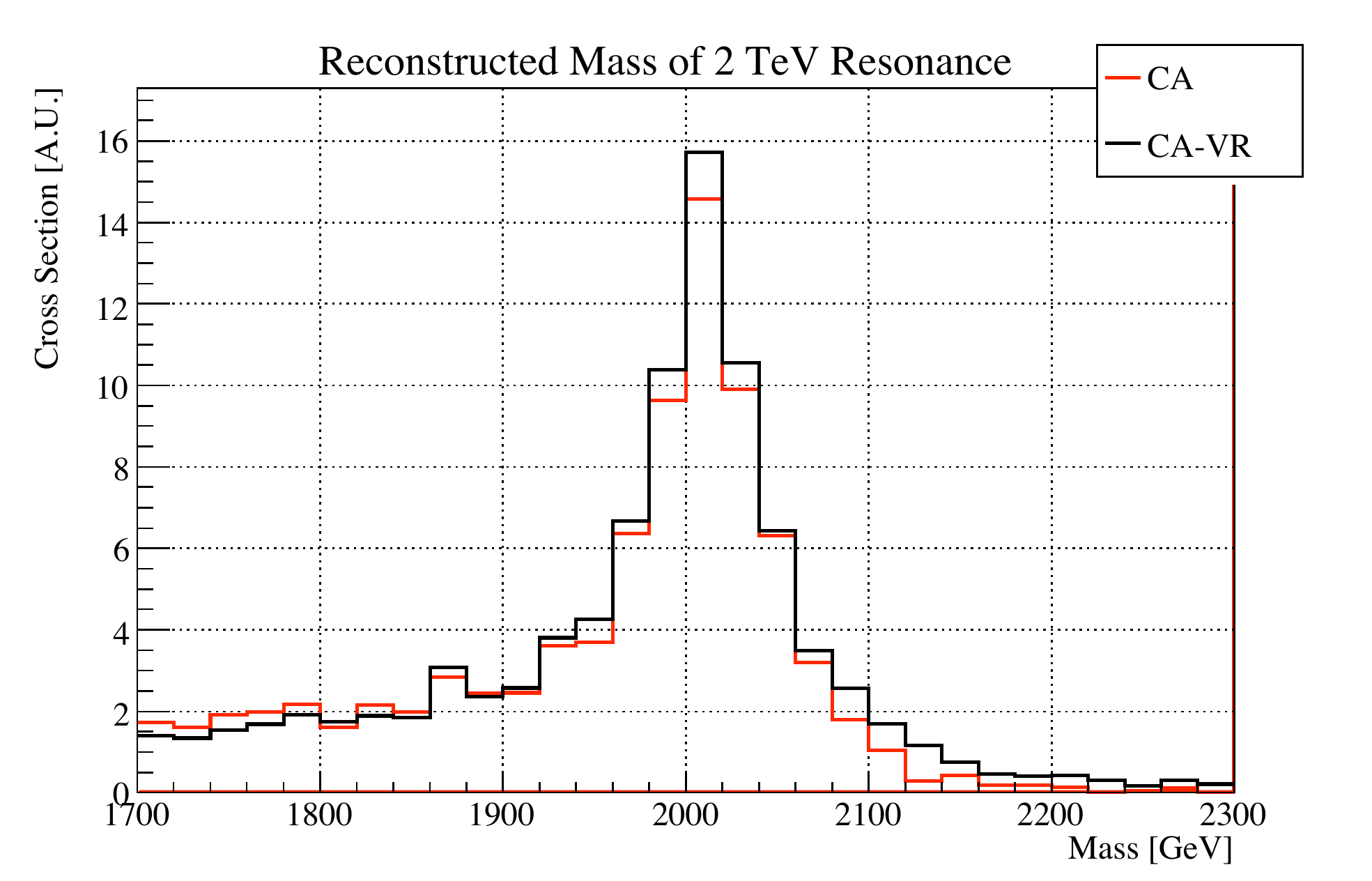}
\includegraphics[scale=0.35]{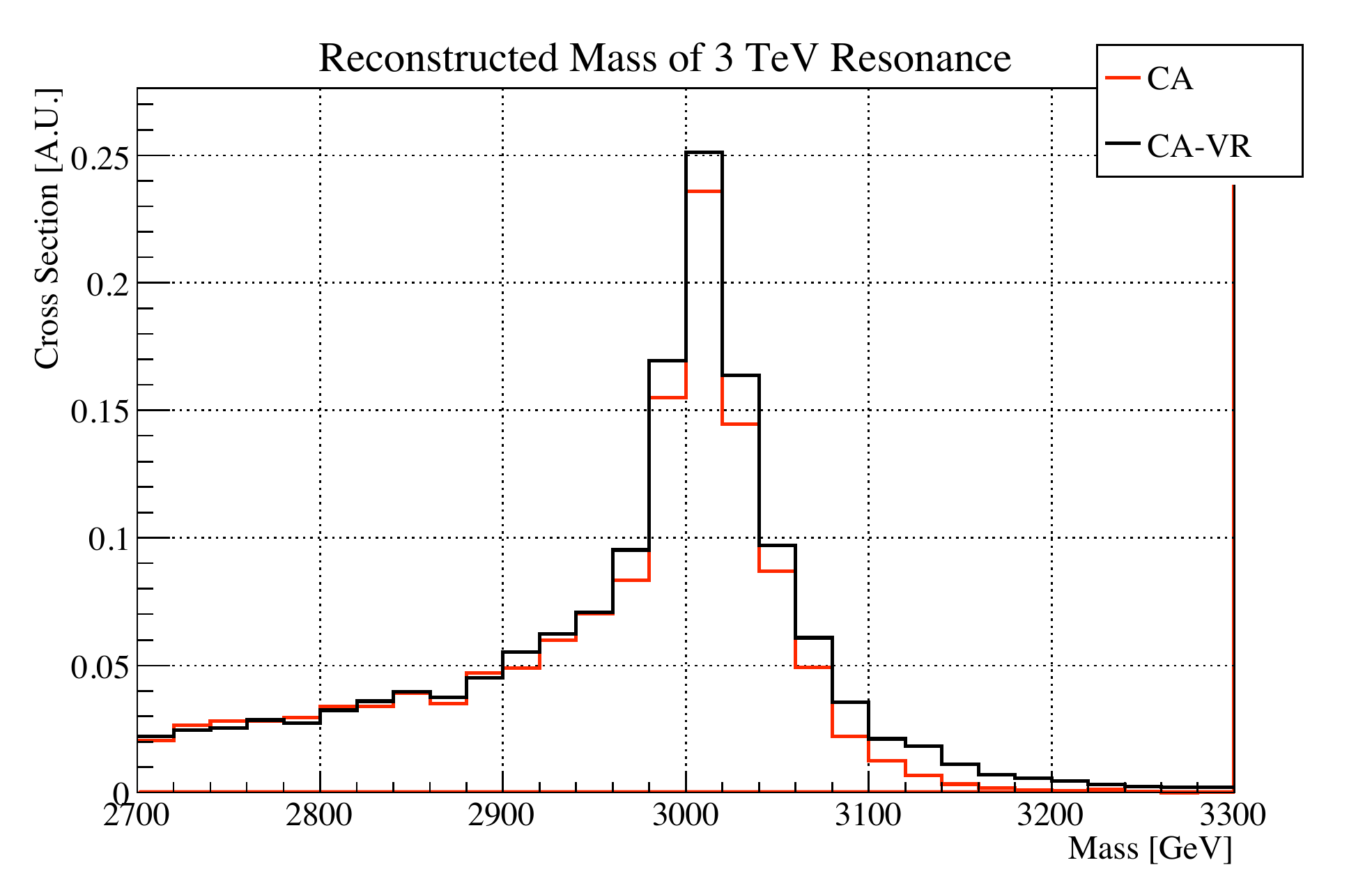}
\caption{\label{fig:resnobgCA} Same as Figure~\ref{fig:resnobgAKT} but comparing the CA and CA-VR algorithms.}
}

The resonance invariant mass plots from this analysis can be seen in Figures~\ref{fig:resnobgAKT} and \ref{fig:resnobgCA}.  The results indicate a uniform improvement in going from the original algorithms to their VR variants: the reconstructed resonances are narrower and taller.  The relatively large cone sizes found in the optimization should not be troubling as similar results were found in Ref.~\cite{Cacciari:2008gd}. The optimized choices of $\rho$ displayed in Table~\ref{tableresnobg} also make intuitive sense.  We expect most of these jets to have a $p_T$ slightly below half the resonance mass, so for $\rho\approx m_X/2$, $R_{\rm eff}$ will be $\mathcal{O}(1)$, close to the optimized value found for the fixed $\Delta R$ algorithms.  Note that the values of $\rho$ are consistent with the discussion in Appendix~\ref{app:applic}.

It is interesting that the improvements offered by the VR algorithms are greatest for small resonance masses.  The reason is that signal degradation can come from out-of-cone corrections and contamination from the underlying event (including ISR).  For larger resonance masses with high $p_T$ jets, underlying event corrections are proportionally smaller, so one need only make the cone size sufficiently large to capture the bulk of the resonance signal.  For smaller resonance masses with low $p_T$ jets, the underlying event corrections are more important, and the VR algorithms do a better job balancing the need for large cones to capture the signal against the need for suppressing contamination.  

As a side note, the AKT algorithm does perform better than the CA algorithm, but by only 1--5\% in the cases considered.  Thus, the improvement shown by the VR algorithms is greater than the difference between the two fixed cone algorithms.

\subsection{Longer Decay Chains}
The scenario considered above involved a simple event topology and relatively large jet cones.  Moreover, the jets were only boosted along the beam axis, so we expect them to be approximately circular in $(\eta,\phi)$.  Here we test a more complex scenario where the final state is more crowded (requiring smaller cones), and the final state jets can be boosted along a transverse axis.  As discussed before, when jets are produced in a transversely boosted frame we expect the VR algorithms to capture their scaling, but not their exact shape.  Here we will see that by accounting only for this scaling we are able to realize significant performance improvements.

We use two color-octet scalars, $X$ and $Y$, again with negligible width, of mass $m_X=3~\mathrm{TeV}$ and $m_Y=500~\mathrm{GeV}$, decaying to jets via $gg\rightarrow X\rightarrow YY\rightarrow gggg$. After clustering the jets, we optimize the algorithm parameters to maximize the number of events where two pairs of jets each reconstruct $m_Y$ within $25~\mathrm{GeV}$ of its true value, and all four jets reconstruct $m_X$ within $50~\mathrm{GeV}$.  Here we limit the maximum cone size to $R_{\rm max} = 1.0$.  The results are shown in Table~\ref{tablefj}.  We again see a universal improvement in reconstruction.  The reconstructed distributions for $m_X$ and $m_Y$ can be seen in Figures~\ref{fig:fjheavy} and \ref{fig:fjlight}, respectively.

\TABLE[h]{
\parbox{\textwidth}{
\begin{center}
\begin{tabular}{c|c}
\hline 
Algorithm & $X \rightarrow YY$\\
\hline
\hline
AKT $\rightarrow$ AKT-VR      & $15\%$ $(0.7,450)$\\
CA $\rightarrow$ CA-VR            & $23\%$ $(0.6,450)$\\
\hline
\end{tabular}
\end{center}
}
\caption{Percentage increase in the number of reconstructed events for the process $gg\rightarrow X\rightarrow YY\rightarrow gggg$.  We insist that pairs of jets reconstruct $m_Y \pm 25~\mathrm{GeV}$, and that four jets reconstruct $m_X \pm 50~\mathrm{GeV}$.  The numbers in parenthesis are the optimized parameters for the original and VR variant ($R_0$ and $\rho$, with $\rho$ in GeV) respectively.}
\label{tablefj}
}

\FIGURE[tp]{
\includegraphics[scale=0.35]{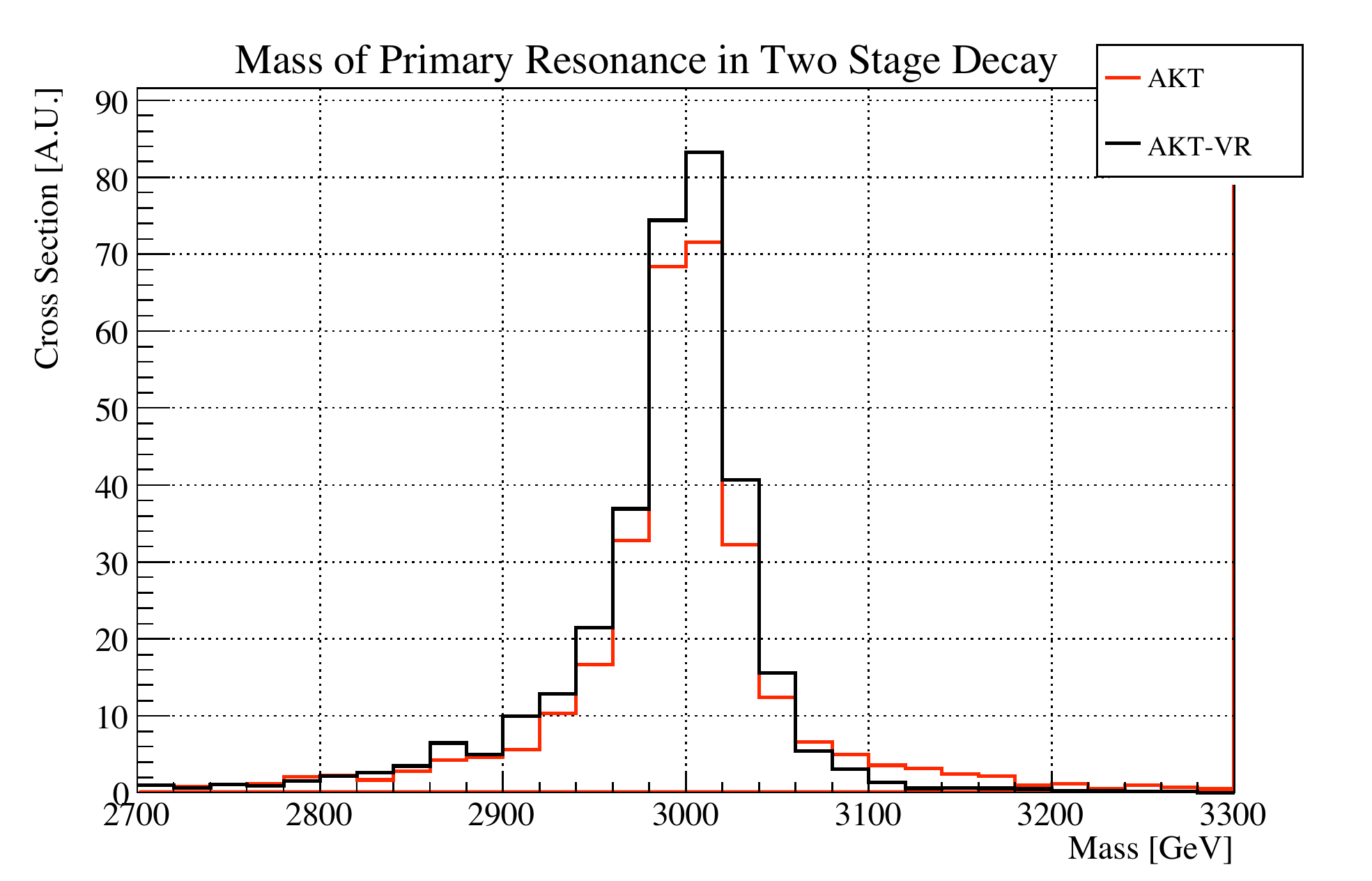}
\includegraphics[scale=0.35]{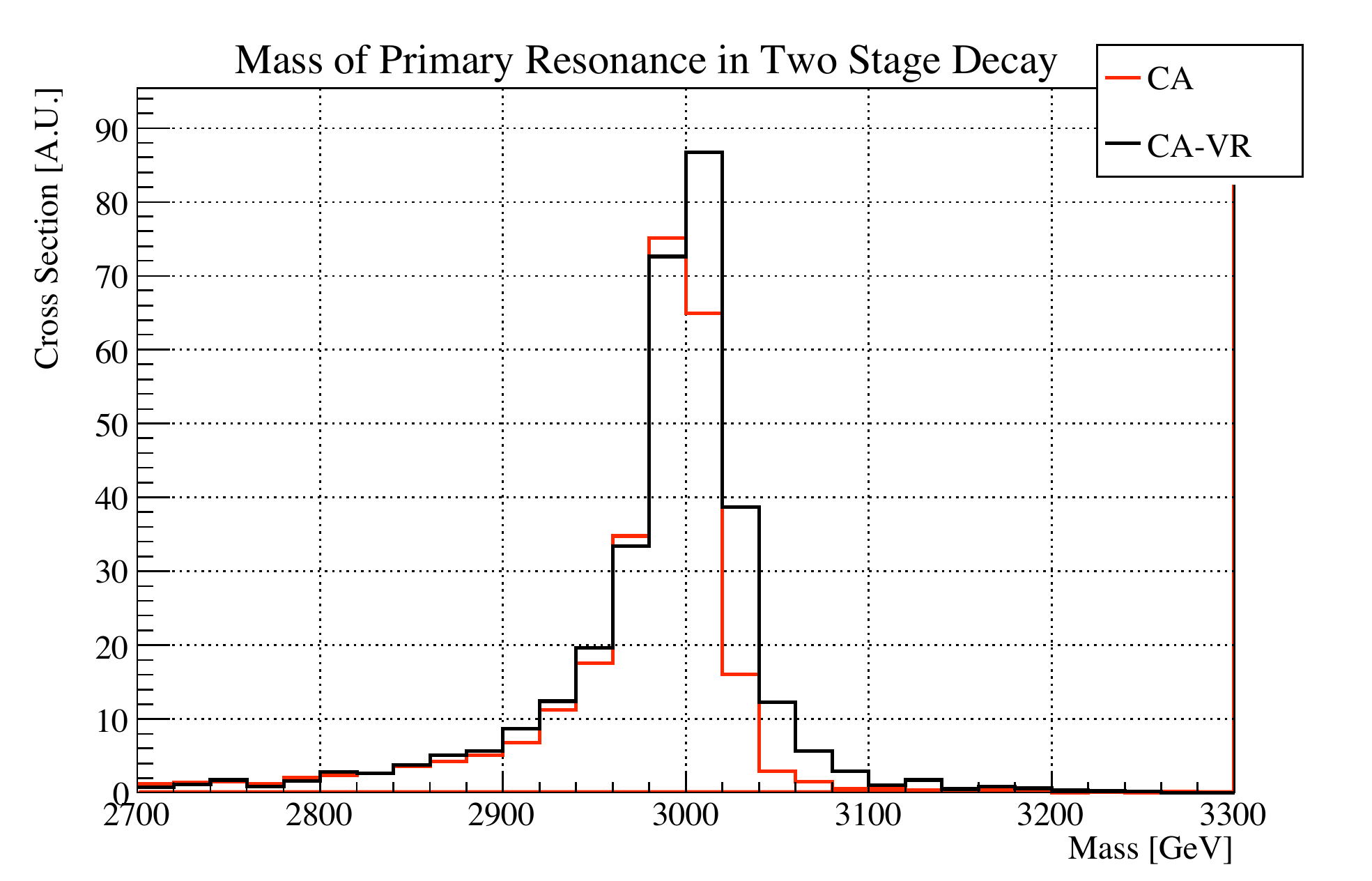}
\caption{\label{fig:fjheavy}Invariant mass distributions of the heavy resonance $X$ for the $X \rightarrow YY \rightarrow gggg$ cascade decay scenario using the optimized parameters in Table~\ref{tablefj}.   The distributions have the same normalization, and the $y$-axis is in arbitrary units (A.U.).  These plots are made after insisting that pairs of jets reconstruct the lighter resonance mass $m_Y$ within $25~\mathrm{GeV}$.}
}

\FIGURE[tp]{
\includegraphics[scale=0.35]{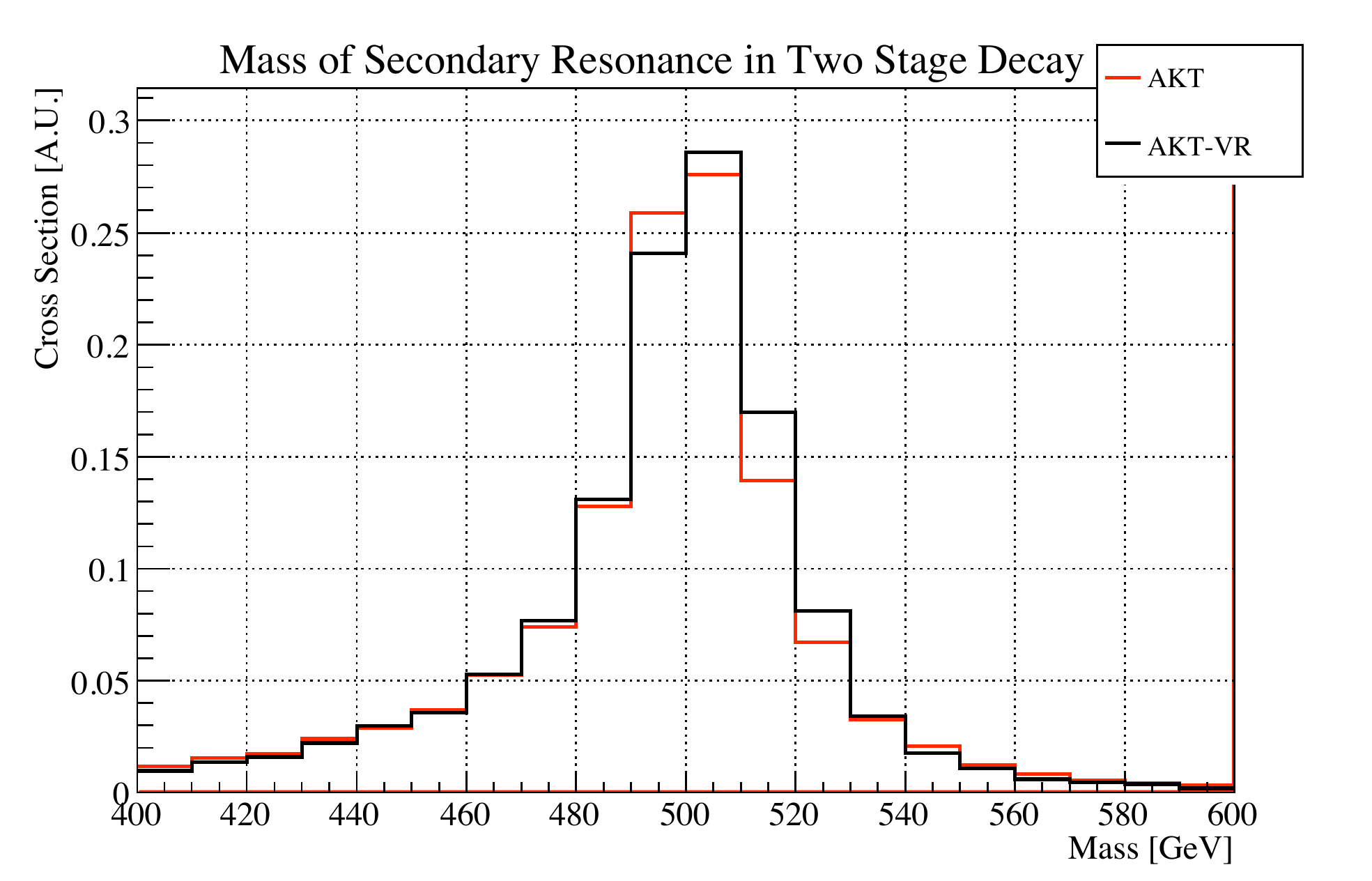}
\includegraphics[scale=0.35]{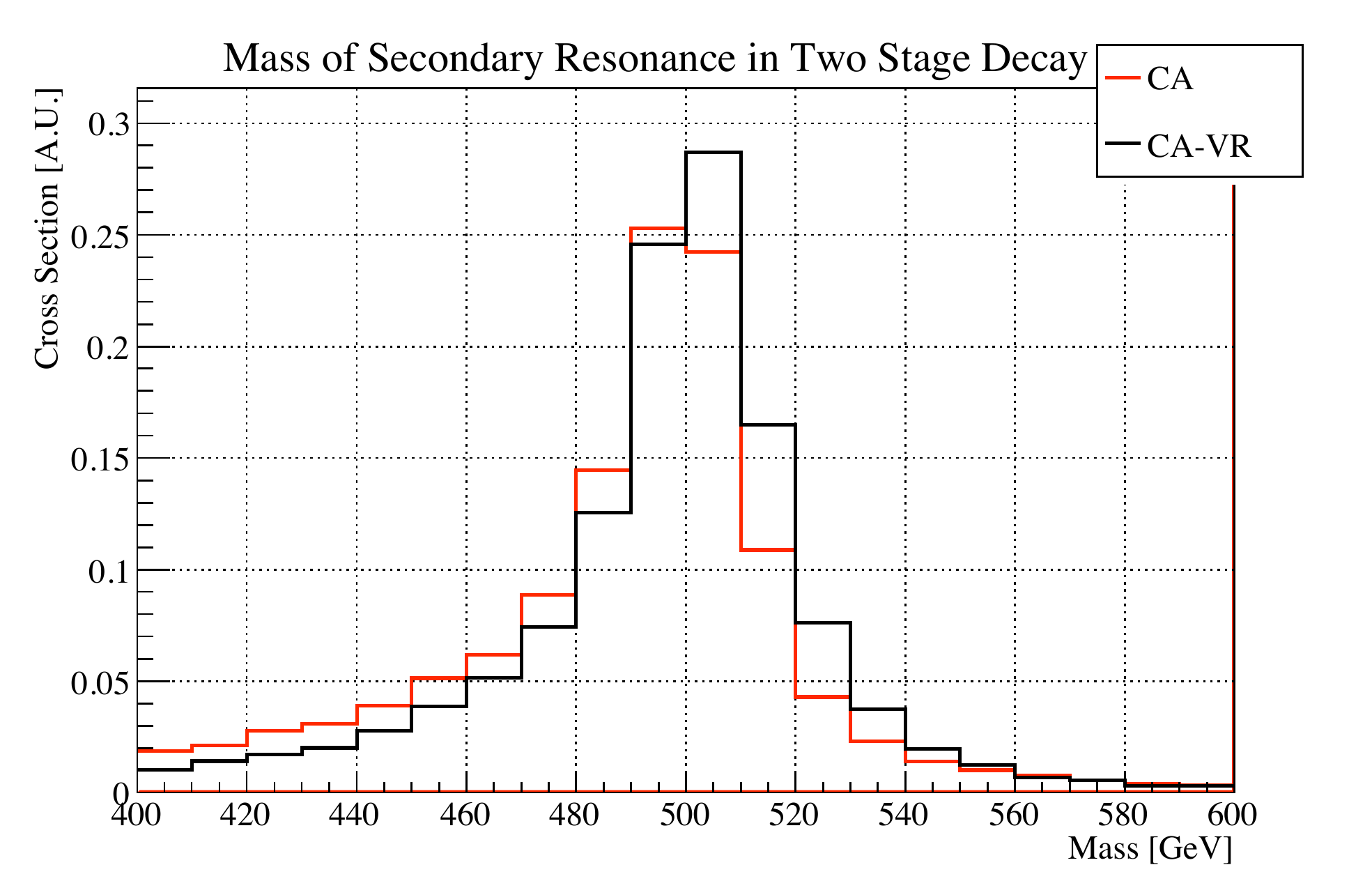}
\caption{\label{fig:fjlight} Same as Figure~\ref{fig:fjheavy} but for the lighter resonance $Y$.}
}

\FIGURE[tp]{
\includegraphics[scale=0.35]{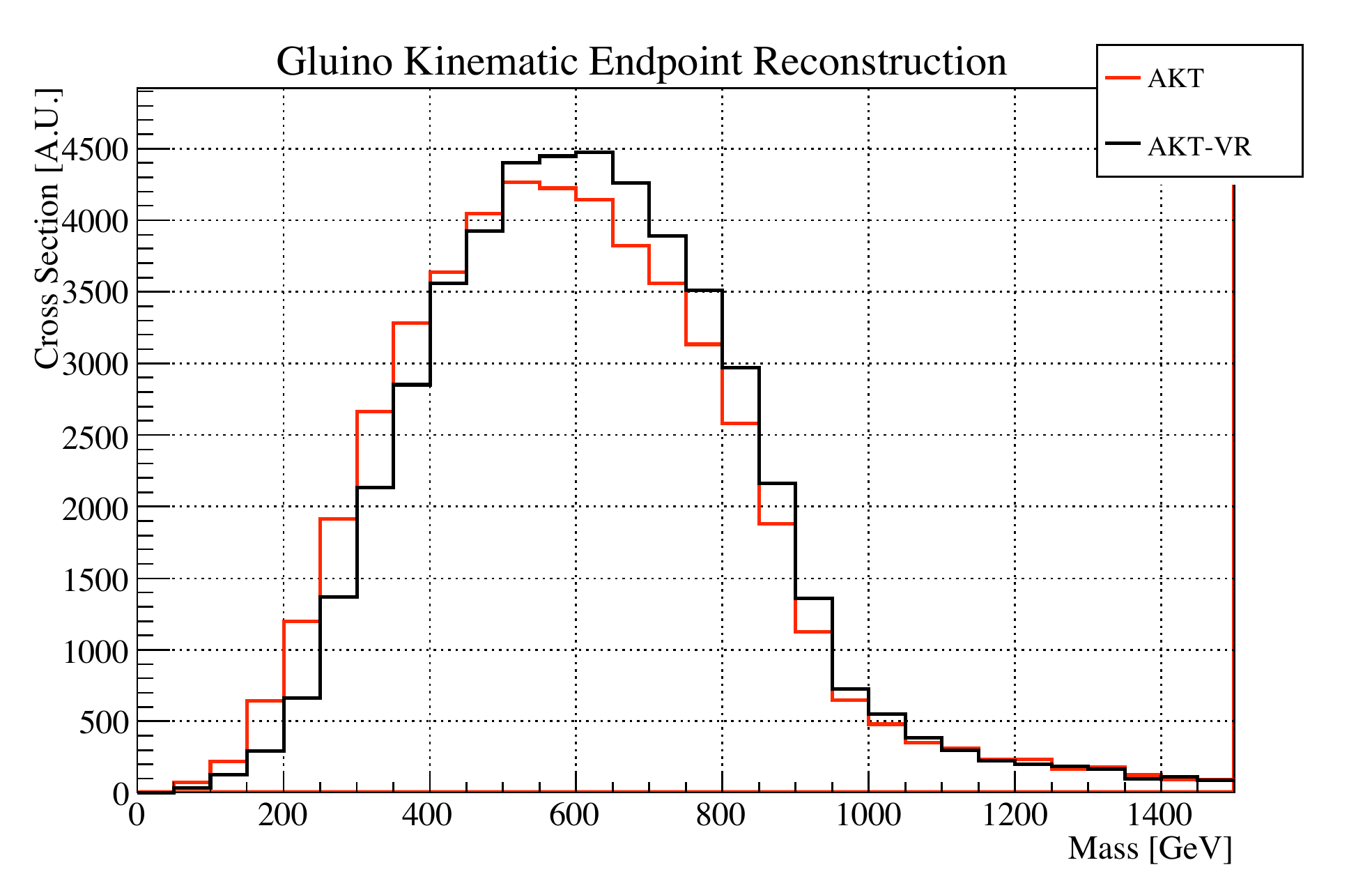}
\includegraphics[scale=0.35]{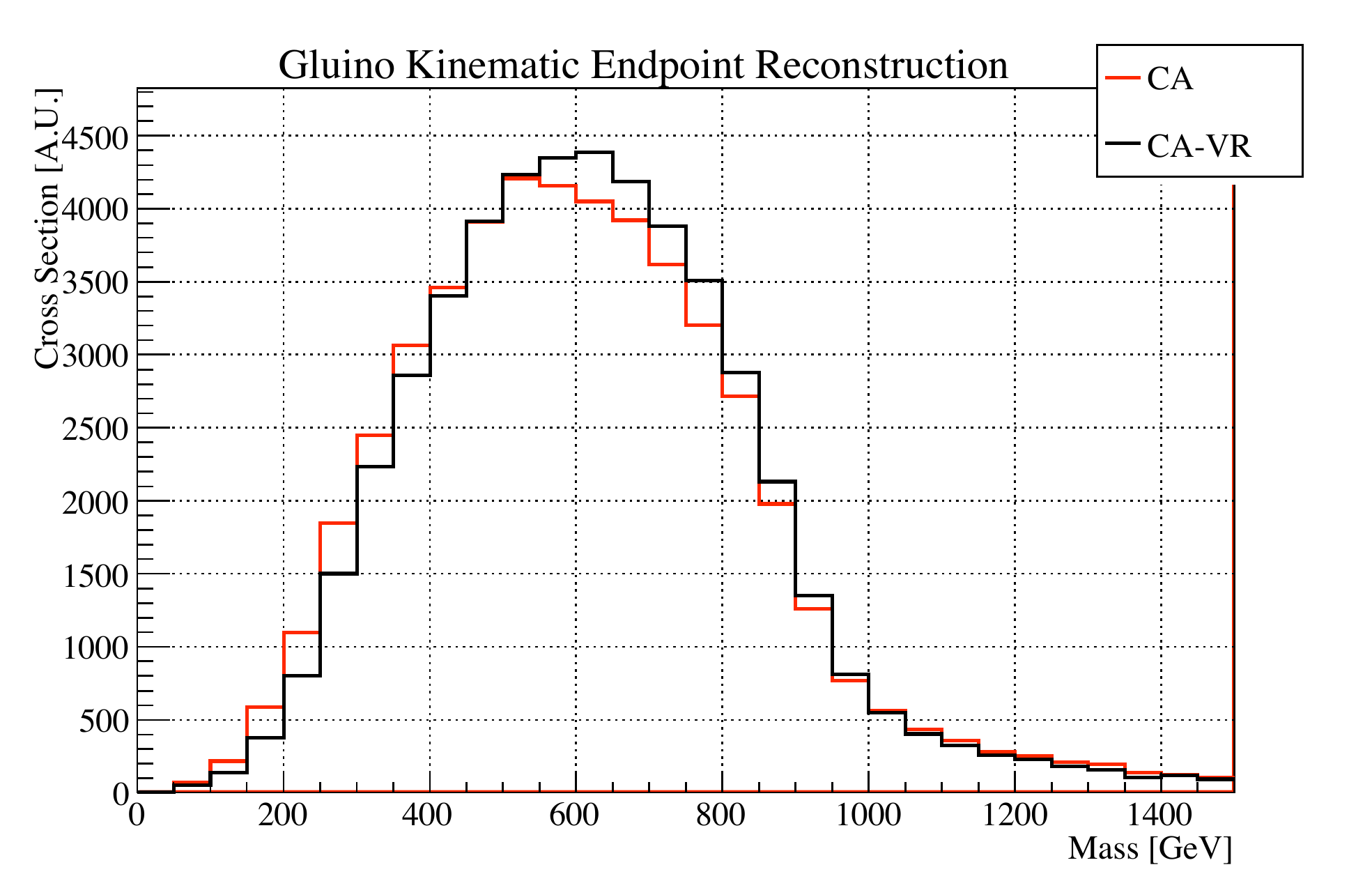}
\caption{\label{fig:susy} Invariant mass distributions of dijets in three-body gluino decay for the optimized parameters in Table~\ref{tablesusy}.   The distributions have the same normalization, and the $y$-axis is in arbitrary units (A.U.).  For the chosen spectrum we expect to see an endpoint at $900~\mathrm{GeV}$.  It can be seen that the VR algorithms fall more sharply near this endpoint than the fixed cone algorithms.}
}

\subsection{Three-Body Gluino Decay}

As we remarked in Section~\ref{sec:vralgs}, we expect the VR algorithms to improve jet reconstruction as long as there is some reference frame in which the jets have the same energy and opening angle.  Here, we demonstrate a useful application of this effect with a gluino decaying to a neutralino lightest supersymmetric particle (LSP), $\tilde{g} \rightarrow q +\bar{q} + \tilde{\chi}^0_1$.  If the intermediate squark is on-shell, then we expect no such reference frame, as the ``upper'' and ``lower'' jets have different preferred mother frames.    However, if the squark is off-shell, then the process is a three-body decay, and the gluino rest frame is a preferred frame for both jets.  Moreover, we expect VR-symmetry to be enhanced near the kinematic endpoint where in the gluino rest frame, the two jets are back-to-back while the LSP is at rest.  Since gluinos are typically produced with small transverse boosts in the lab frame, we do not expect large distortions of the jet shape.

To test our algorithm, we consider the associated production of gluinos with a neutralino LSP via $f\bar{f}\rightarrow \tilde{g} + \tilde{\chi}^0_1$ and use a spectrum where $m_{\tilde{g}}=1~\mathrm{TeV}$ and $m_{\tilde{\chi}_1^0}=100~\mathrm{GeV}$.  We choose the associated production channel for demonstration purposes to eliminate the combinatorial confusion present in gluino pair production.   To  ensure a three-body decay, we have lifted the squarks out of the spectrum by placing them at $5~\mathrm{TeV}$.  With this spectrum, we expect to see an endpoint in the dijet invariant mass distribution at $900~\mathrm{GeV}$.  The more accurate a jet algorithm, the better it can reconstruct this endpoint.  Therefore, we define the measure of reconstruction performance to be the  \emph{difference} in the number of events reconstructed in $100~\mathrm{GeV}$ windows above and below $900~\mathrm{GeV}$.   We have optimized this measure with $R_{\max} = 1.5$.   The improvement of the VR variants is shown in Table~\ref{tablesusy} and Figure~\ref{fig:susy}.  

\TABLE[h]{
\parbox{\textwidth}{
\begin{center}
\begin{tabular}{c|c}
\hline 
Algorithm & $\tilde{g} \rightarrow q +\bar{q} + \tilde{\chi}^0_1$ \\
\hline
\hline
AKT $\rightarrow$ AKT-VR      & $14\%$ $(1.2,600)$\\
CA $\rightarrow$ CA-VR           & $7\%$ $(1.3,650)$\\
\hline
\end{tabular}
\end{center}
}
\caption{Improvement, measured as the difference in the number of events reconstructed in $100~\mathrm{GeV}$ bins on either side of the endpoint, in using VR algorithms to reconstruct three-body gluino decays.  The numbers in parenthesis are the optimized parameters for the original and VR variant ($R_0$ and $\rho$, with $\rho$ in GeV) respectively.}
\label{tablesusy}
}


\section{Resonance Decays With Background}
\label{sec:rdwb}

In the previous section, we showed that VR algorithms generically lead to improvements in signal reconstruction for events meeting our VR-symmetric criteria.  It is natural to wonder how the VR algorithms will handle background events.  The VR algorithms have a dimensionful parameter $\rho$, unlike their fixed cone counterparts, so some shaping of the background might take place. However, we will see that this is not the case when one imposes reasonable jet quality cuts.\footnote{These quality cuts were not used in our signal-only analysis.} 

\FIGURE[tp]{
\includegraphics[scale=0.35]{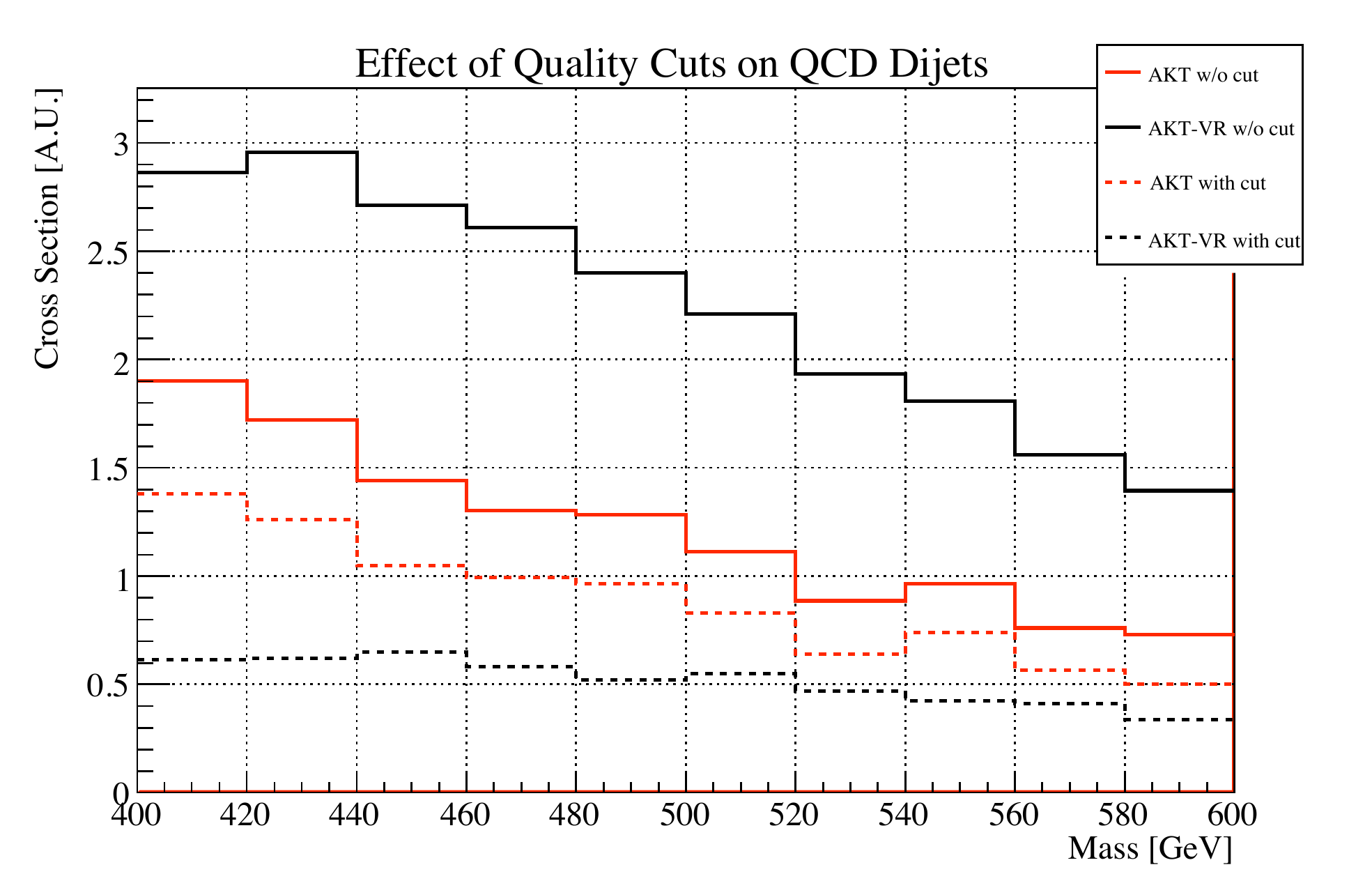}
\includegraphics[scale=0.35]{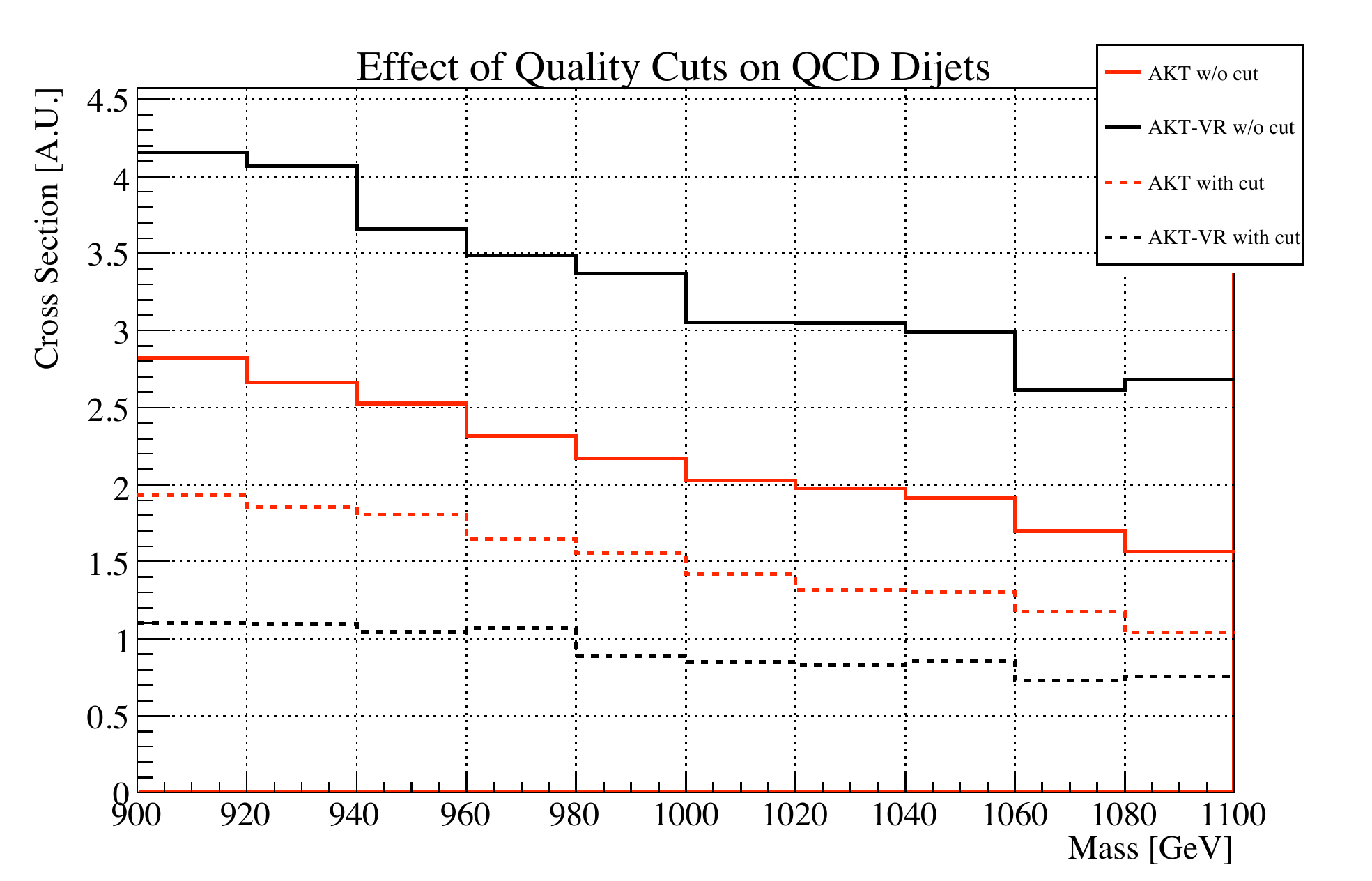}
\\
\includegraphics[scale=0.35]{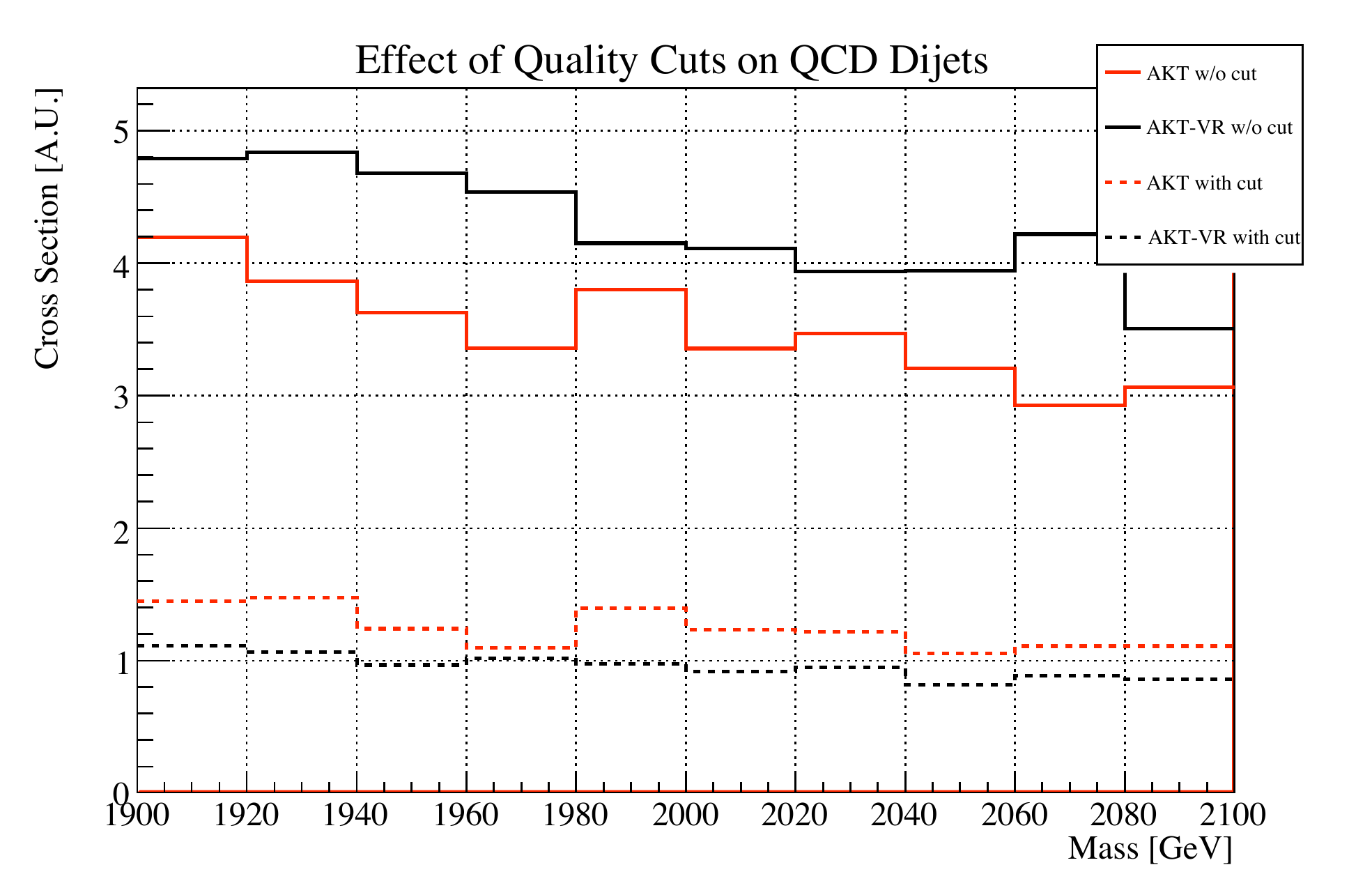}
\includegraphics[scale=0.35]{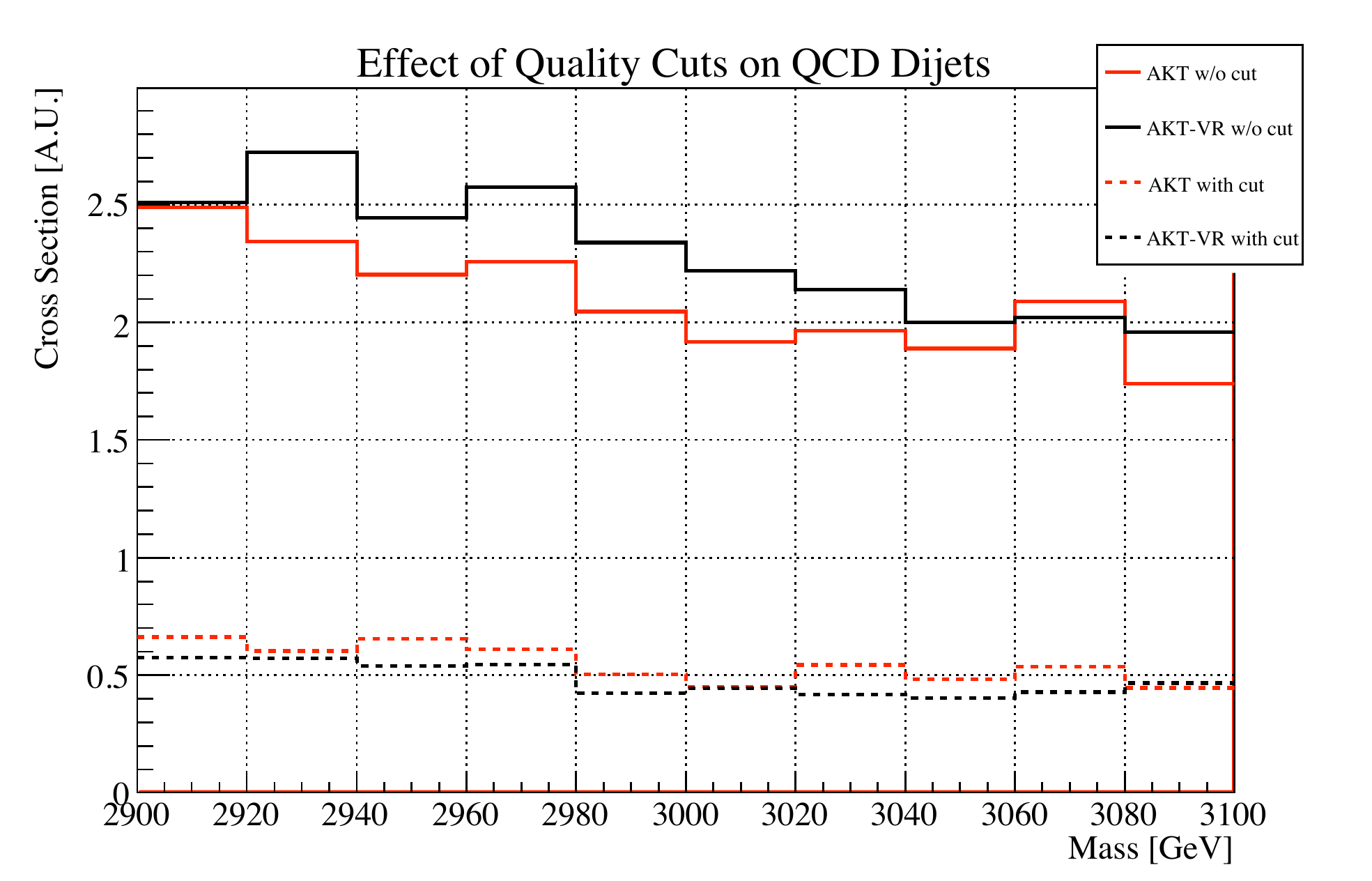}

\caption{\label{fig:bkgd} Invariant mass distributions of the QCD dijet background for the AKT and AKT-VR algorithms, plotted in windows corresponding to the four fiducial resonances.  The distributions have the same normalization, and the $y$-axis is in arbitrary units (A.U.).  The solid (dotted) lines show the background without (with) quality cuts.  The red lines show the original algorithm, and the black lines show the VR modification.  The results are qualitatively similar for CA and CA-VR.}
}

\FIGURE[tp]{
\includegraphics[scale=0.35]{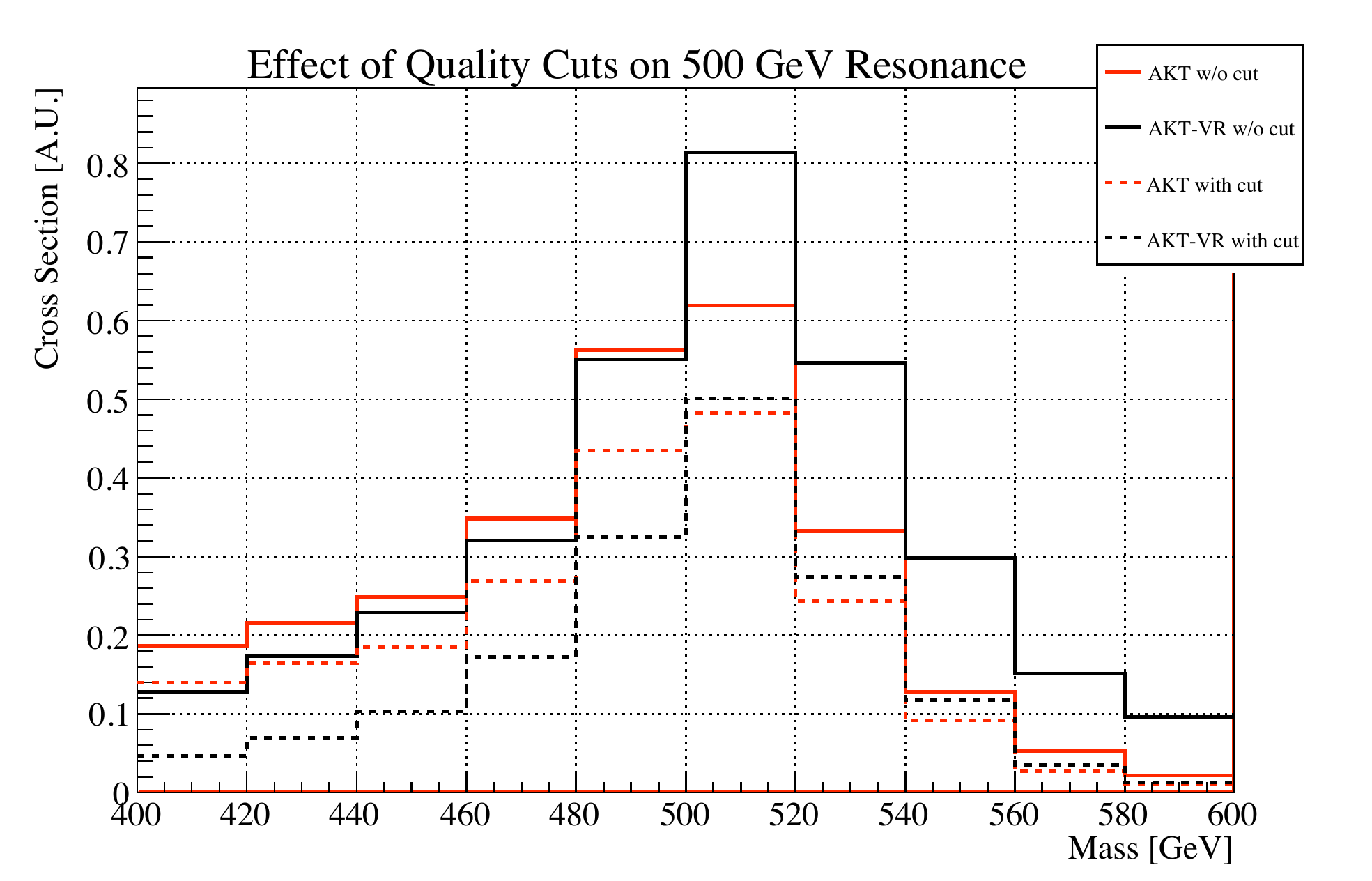}
\includegraphics[scale=0.35]{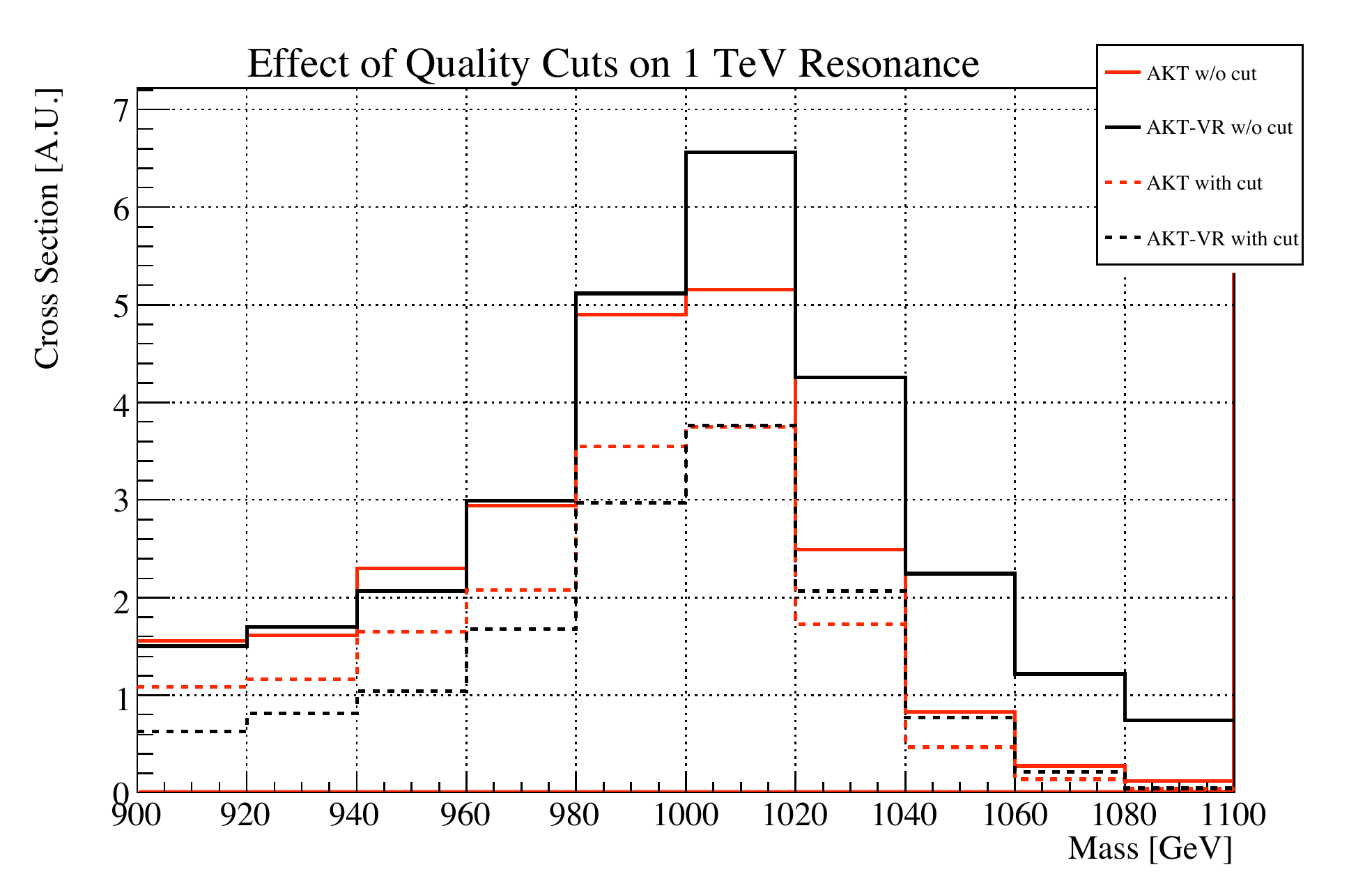}
\\
\includegraphics[scale=0.35]{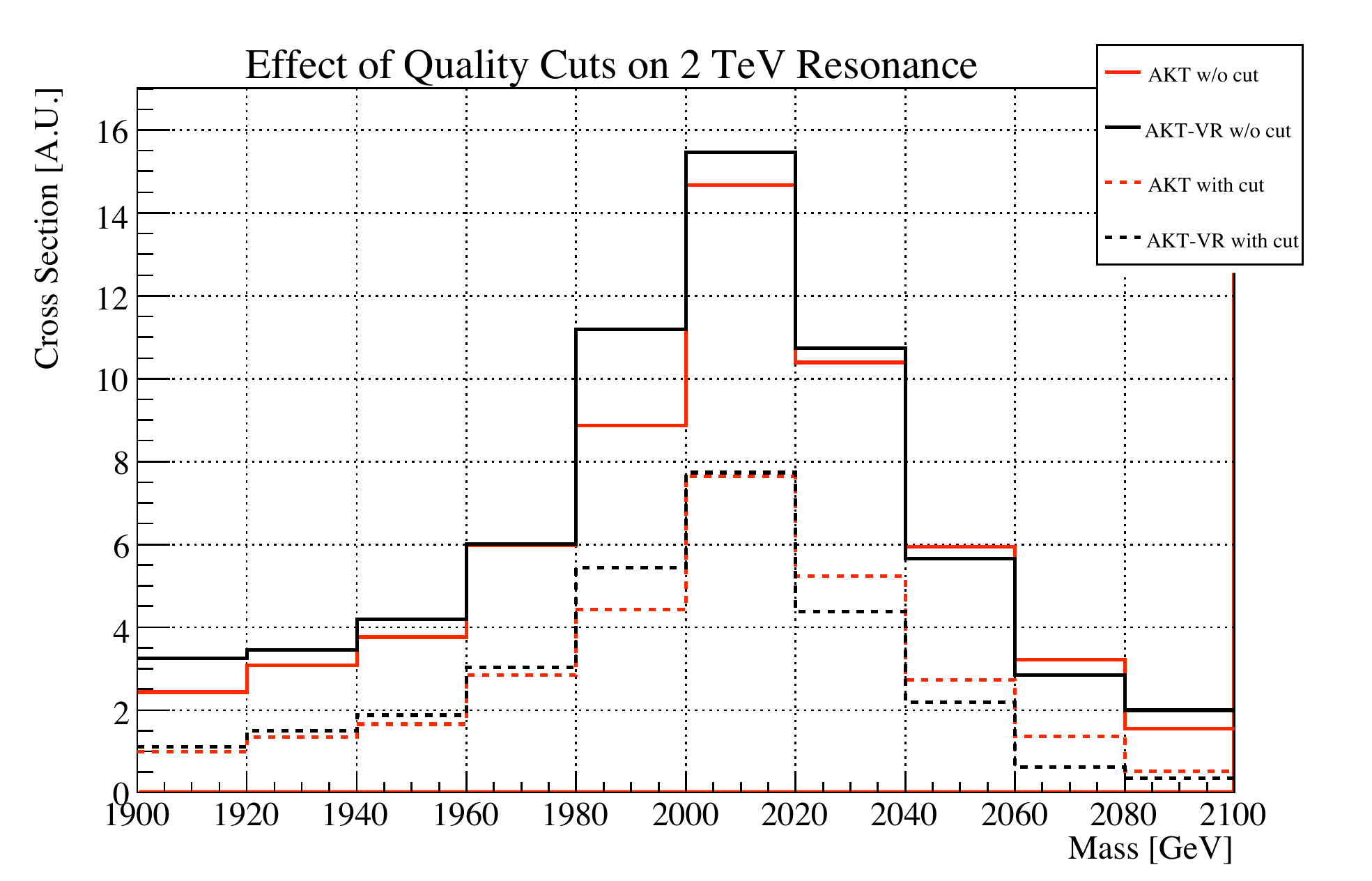}
\includegraphics[scale=0.35]{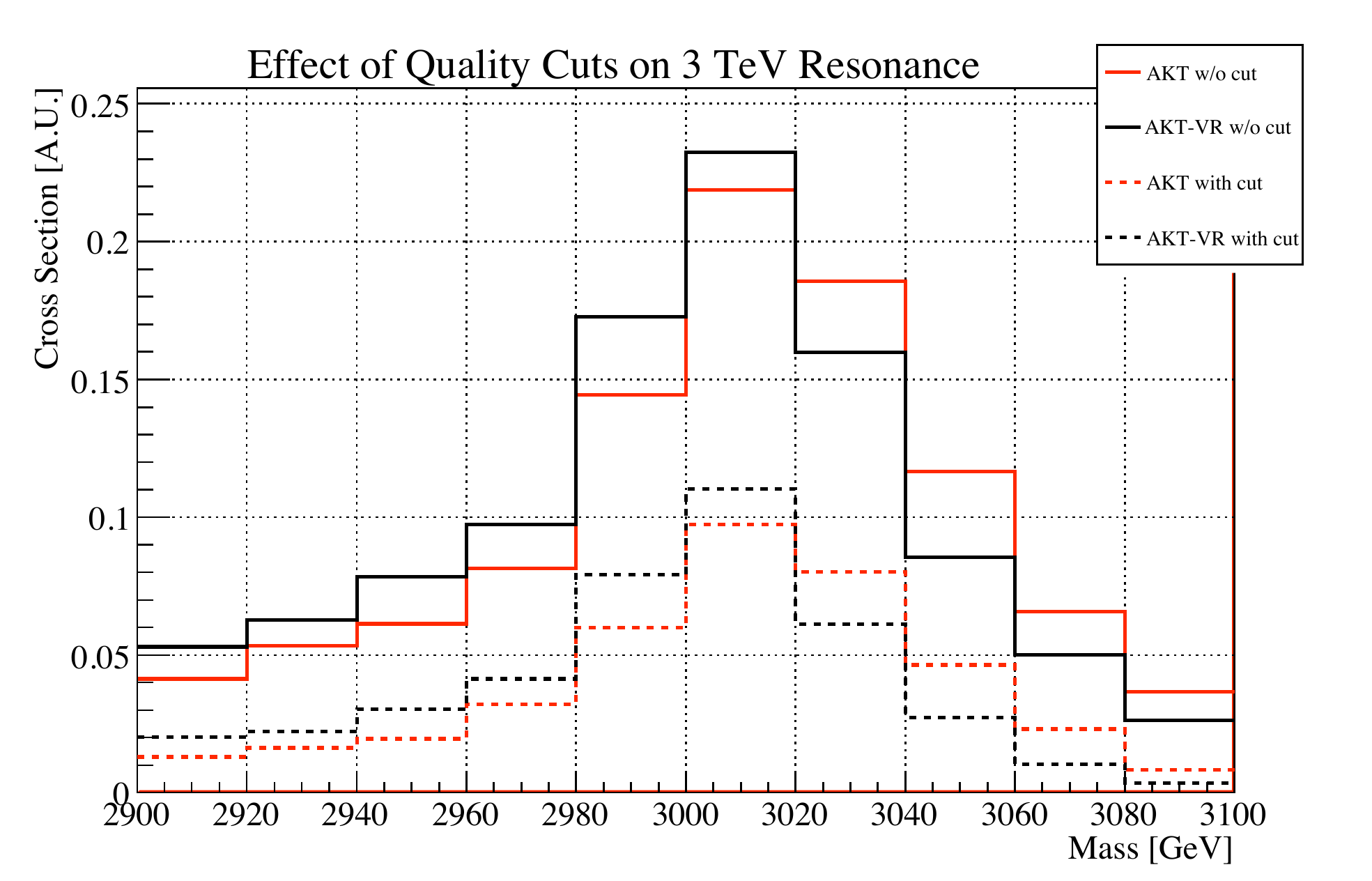}
\caption{\label{fig:sigwwocuts} Same as Figure~\ref{fig:bkgd} but for the resonance signals.}
}

To understand the effect of the VR algorithms on background, we consider the color-octet resonances previously described in Section~\ref{subsec:resonancenobackground} on top of a background of QCD dijets.  We optimize our algorithm parameters to maximize signal significance, defined as $S/\sqrt{B}$, where $S$ and $B$ are the signal and background cross sections, respectively, in a given mass window.  In addition, we remove the enormous low-$p_T$ QCD background by requiring the two hardest jets each satisfy $p_T > m_{\rm res.}/4$.

Let us first consider the dijet background without any kind of quality cut, as shown by the solid-lined histograms in Figure~\ref{fig:bkgd}.  Here we see significant background shaping by the VR algorithms:  they are taking low $p_T$ jets from events with small dijet invariant mass, assigning the jets large cones, and thus pushing the events to a higher dijet invariant mass compared to the fixed cone algorithms.  Because the QCD background exhibits a steeply falling distribution in invariant mass, this increases the normalization of the background in the mass windows shown in Figure~\ref{fig:bkgd}.  However, this problem is not particular to the VR algorithms, and fixed cone algorithms with large $R_0$ will inevitably increase the background to an analysis by having cones that swallow soft particles toward their edges.

Such contamination can be effectively controlled by jet quality cuts, which can take many different forms.  For example, we could place cuts limiting jets to the central region of the detector.  Another possibility would involve systematically subtracting a four-momenta related to the ``catchment area'' of a jet~\cite{Cacciari:2008gn}.  Here, we explore another approach by requiring the energy- and $p_T$-weighted jet centers approximately coincide.  In practice, this means imposing
\begin{equation}
\label{eq:jetquality}
\Delta R(\vec{P}_E,\vec{P}_{p_T}) < \delta,
\end{equation}
where $\vec{P}_E$ and $\vec{P}_{p_T}$ are the energy- and $p_T$-weighted jet centers defined by
\begin{equation}
\vec{P}_E = \sum_{i}E_i \, \hat{p}_i, \qquad \vec{P}_{p_T} = \sum_{i}p_{Ti} \, \hat{p}_i, 
\end{equation}
and $\hat{p}_i$ is the massless four-vector of unit length corresponding to the $i$th calorimeter cell.  

It is reasonable to impose the jet quality standard of \Eq{eq:jetquality} because, as discussed in Ref.~\cite{Dasgupta:2007wa}, we expect jets from a massless parton to have a small, hard central core.  For our analysis we choose $\delta = (0.04,0.025,0.015,0.01)$ for $m_X = (500~\mathrm{GeV},1~\mathrm{TeV},2~\mathrm{TeV},3~\mathrm{TeV})$.  The result of imposing such a cut on the background is shown in Figure~\ref{fig:bkgd}.  These $\delta$ values were chosen by estimating the characteristic size of $\Delta R(\vec{P}_E,\vec{P}_{p_T})$ for each mass window.  We see that this helps significantly in reducing background.  If such a quality cut reduced the signal as much as the background it would be of no use, but as can be seen in Figure~\ref{fig:sigwwocuts}, these cuts have a milder effect on the signal than on background, since the signal jets do have hard central cores.  One expects that the $\delta$ parameter could be optimized to increase the signal yield, but we will not pursue this possibility here.

Now we employ the quality cuts and present the improvement in $S/\sqrt{B}$ significance by using the VR algorithms.   We consider a narrow mass window ($m_X \pm25~\mathrm{GeV}$) around the resonance as in the signal-only study from Section~\ref{subsec:resonancenobackground}.  The results are shown in Table~\ref{tablesigbgfixed}, and we again see a universal improvement in going to VR.  The jet parameters are comparable to the ones seen in the signal-only study of Table~\ref{tableresnobg}.

\TABLE[h]{
\parbox{\textwidth}{
\begin{center}
\begin{tabular}{c|cccc}
\hline 
Algorithm & $500~\mathrm{GeV}$ & $1~\mathrm{TeV}$ & $2~\mathrm{TeV}$ & $3~\mathrm{TeV}$\\
\hline
\hline
AKT $\rightarrow$ AKT-VR    & $19\%$ $(0.9,225)$ & $23\%$ $(0.9,450)$ & $18\%$ $(1.2,1000)$& $21\%$ $(1.3,1500)$\\
CA $\rightarrow$ CA-VR           & $10\%$ $(0.8,200)$& $18\%$ $(0.9,450)$& $15\%$ $(1.2,1000)$& $18\%$ $(1.3,1500)$\\
\hline
\end{tabular}
\end{center}
}
\caption{Percentage increase in $S/\sqrt{B}$ significance for resonance reconstruction over background within a \emph{fixed} mass window $m_X\pm25~\mathrm{GeV}$.  The numbers in parenthesis are the optimized parameters for the original and VR variant ($R_0$ and $\rho$, with $\rho$ in GeV) respectively.}
\label{tablesigbgfixed}
}

While accurate mass reconstruction is important for measuring the properties of a hadronic resonance, for discovery of a resonance one simply wants to increase $S/\sqrt{B}$.  Thus, we also consider a floating mass window, of variable size around $m_X$, chosen to maximize $S/\sqrt{B}$ for each jet parameter.   Results for this study are shown in Table~\ref{tablesigbgvar}.  Here the improvement in going to VR is somewhat less than in the fixed window case, but still notable.

\TABLE[h]{
\parbox{\textwidth}{
\begin{center}
\begin{tabular}{c|cccc}
\hline 
Algorithm & $500~\mathrm{GeV}$ & $1~\mathrm{TeV}$ & $2~\mathrm{TeV}$ & $3~\mathrm{TeV}$\\
\hline
\hline
AKT $\rightarrow$ AKT-VR    & $10\%$ $(0.6,225)$ & $21\%$ $(0.9,500)$ & $14\%$ $(1.2,1000)$& $8\%$ $(1.4,1600)$\\
CA $\rightarrow$ CA-VR          & $3\%$ $(0.6,200)$& $16\%$ $(0.8,450)$ &$11\%$ $(1.2,1000)$& $13\%$ $(1.4,1700)$\\
\hline
\end{tabular}
\end{center}
}
\caption{Percentage increase in $S/\sqrt{B}$ significance for resonance reconstruction over background in a \emph{variable} mass window.  The numbers in parenthesis are the optimized parameters for the original and VR variant ($R_0$ and $\rho$, with $\rho$ in GeV) respectively.}
\label{tablesigbgvar}
}


\section{Conclusion}
\label{sec:conc}

We have constructed a new class of jet algorithms in which jet radii $\Delta R$ become functions of jet $p_T$.  Although we have only explored the simplest new algorithm in this class, the rules described in Appendix~\ref{app:irsafety} should provide a well-defined sandbox in which more complicated variants can be explored.  Surely undiscovered algorithms exist within this framework that can further improve jet-based analyses at hadron colliders.

In this paper, we focused on the simplest variable $\Delta R$ algorithm, denoted VR, in which $\Delta R$ scales as $1/p_T$.  Remarkably, this scaling captures the physical size of jets for many diverse processes.  To test this algorithm, we developed a sequential-recombination implementation using the \texttt{FastJet} \cite{Cacciari:2005hq,Cacciari:Fastjet} package, extending the Cambridge-Aachen \cite{Dokshitzer:1997in,Wobisch:1998wt} and anti-$k_T$ \cite{Cacciari:2008gp} algorithms to VR variants.  In the analysis of single/multiple resonance decays we routinely saw 10 -- 20\% increases in signal efficiency when using the VR algorithms.  A similar improvement appears in reconstructing the kinematic endpoint in three-body gluino decay.

In order to use the VR algorithms in the presence of a significant continuum jet background, we needed to impose jet quality cuts.  We emphasize that these cuts are useful in and of themselves, and have strong physical motivation, since jets from hard partons have different substructure compared to jets formed from soft radiation.  After quality cuts are imposed, the VR algorithms outperform their fixed cone  cousins by 10 -- 20\% in statistical significance.

Given our success in developing a jet algorithm for VR-symmetric event topologies, one could imagine developing more powerful ``designer'' jet algorithms to improve signal acceptance or background rejection for specific physics scenarios.   The original anti-$k_T$ algorithm emphasizes the utility of recursive jet algorithms, since it returns idealized cone jets in an infrared/collinear safe manner.  The AKT-VR variant emphasizes the flexibility of recursive jet algorithms to adapt to different event topologies.  One suspects that by including additional global event data (or even jet substructure \cite{Butterworth:2002tt,Butterworth:2008iy,Thaler:2008ju,Kaplan:2008ie,Almeida:2008yp} data) in the definitions of $d_{ij}$ and $d_{iB}$, one could better extract information about specific hard scattering processes despite the complicated hadronic environment.

\acknowledgments{
Inspiration for this paper came from a KITP conference talk by John Conway in 2006, which questioned the desirability of fixed $\Delta R$ cones.  We thank Gavin Salam for patient assistance with \texttt{FastJet} and for valuable comments on a draft of this paper.   J.T. is supported by the Miller Institute for Basic Research in Science.  L.-T. W. is supported by the National Science Foundation under grant PHY-0756966 and the Department of Energy under grant DE-FG02-90ER40542.}


\appendix

\section{Effective Jet Radii}
\label{app:irsafety}

It is important for jet algorithms to be infrared/collinear safe, since this ensures insensitivity to detector effects and is necessary for meaningful higher order calculations~\cite{Sterman:1977wj}.  Since variable $\Delta R$ algorithms are extensions of the jet parameterization from \Eq{eq:param}, it is easy to check that they inherit infrared/collinear safety from standard recursive jet algorithms.  However, the purpose of variable $\Delta R$ algorithms is to define jets with an effective radius $R_{\rm eff}$, and infrared/collinear safety alone is insufficient to guarantee a reasonable notion of jet radius.   Therefore, we will impose a stronger condition, requiring our algorithms to be ``collinear robust''.

We focus our attention on jet algorithms with $n\leq0$, because these are the only algorithms for which the concept of an effective radius is meaningful.  The reason for this is simple: to talk about an effective radius, the jets must be approximately circular, and the only way to achieve this with a sequential recombination algorithm is to start with a central core and add on to it.  This is achieved for $n\leq 0$ because clustering begins with the hardest four-momenta.\footnote{For the marginal case where $n=0$ (so clustering is by angle), this still works because the hard center of a jet sees a high concentration of radiation at small angular separation.}  The $k_T$ algorithm does not satisfy this condition, since the algorithm clusters from soft objects to hard ones, and the final shape of a $k_T$ jet is only determined in the last few stages of clustering.  As a result, $k_T$ jets assume non-circular shapes, and we are unable to assign these jets a meaningful effective radius.

It is useful to consider three degrees of collinear robustness.  The weakest of these demands that the clustering of two four-momenta is controlled by the effective jet radius of the harder one.  This ensures that the hardest jets in an event have well-defined jet radii, and non-circular effects only appear at lower $p_T$.  Consider two four-vectors $i$ and $j$, with $p_{Ti} > p_{Tj}$.  For these two four-vectors to be clustered together, they must satisfy both $d_{ij} < d_{iB}$ and $d_{ij} < d_{jB}$.  For $n \leq 0$, the first inequality yields
\be
\frac{d_{ij}}{d_{iB}} = \frac{R_{ij}^2}{R_{\rm eff}(p_{Ti})} < 1,
\ee
and $R_{\rm eff}$ defines an effective radius as desired.  To make sure that the second inequality does not affect the clustering and ruin the interpretation  of an effective radius, we require $d_{iB} \leq d_{jB}$, which implies
\be
\label{eq:robustA}
p_{Ti}^n R_{\rm eff}(p_{Ti}) \leq p_{Tj}^n R_{\rm eff}(p_{Tj}). 
\ee
This is the requirement for a minimal degree of collinear robustness, and is satisfied by both the fixed cone and VR ($\Delta R \propto 1/p_T$) algorithms.

A stronger version of collinear robustness is that the jet algorithm should be insensitive the resolution of the calorimeter.  Even if \Eq{eq:robustA} is satisfied, it is possible for the jets formed with a fine calorimeter resolution to be different from the jets formed with a coarser resolution.  This can happen, for example, when $n = -1$ and $R_{\rm eff} \propto p_T$, where the effective jet radius can become comparable to the calorimeter resolution, and a coarse calorimeter can form a single jet out of two four-momenta that would not otherwise be clustered using $R_{\rm eff}$.  To guard against this pathology, one can impose
\be
\label{eq:robustB}
R_{\rm eff}(p_T) \gg R_{\rm calorimeter}
\ee
for all values of $p_T$.   For the VR algorithm, this constraint only applies for very high $p_T$ jets, and in practice is never needed for typical beam energies and calorimeter resolutions.

The strongest version of collinear robustness requires that a jet algorithm should be insensitive to \emph{macroscopic} splittings within the jet radius.  This ensures that reasonable rearrangements of a jet's substructure do not cause the jet to be reconstructed differently.\footnote{By ``reasonable rearrangement'', we have in mind situations where an input four-momenta is broken down into smaller pieces within the original effective jet radius.}  This requires that two four-vectors $p_i$ and $p_j$ should be clustered together if they lie with the effective radius defined by the \emph{sum} four-vector $p_i + p_j$.  Again assuming that $p_{Ti} > p_{Tj}$, the desired condition is that $d_{ij} < d_{iB}$ whenever $R_{ij} < R_{\rm eff}(p_{T(i+j)})$, which implies a ``shrinking cone'' requirement
\be
\label{eq:robustC}
R_{\rm eff}(p_{Ti}) \geq R_{\rm eff}(p_{T(i+j)}).  
\ee
This is satisfied by the VR (and fixed cone) algorithms, since $R_{\rm eff}(p_T)$ is monotonically decreasing (or constant) as $p_T$ increases.   Note that \Eq{eq:robustC} implies both \Eq{eq:robustB} and \Eq{eq:robustA}.  

We emphasize that these requirements of collinear robustness are not necessary for the theoretical consistency of the variable $\Delta R$ algorithms.  From the point of view of perturbative infrared/collinear safety, there is no singularity associated with finite angle splittings.  However, without some version of collinear robustness, the effective jet radii would be difficult to understand, since the jet algorithm would be overly sensitive to the precise four-vectors used in the reconstruction.  The shrinking cone requirement of \Eq{eq:robustC} is the conceptually simplest way to enforce collinear robustness, since it does not require defining an $R_{\rm calorimeter}$ and is independent of the precise choice of $n$.

\section{Valid VR Parameter Range}
\label{app:applic}
Here we discuss the operational range of the jet algorithms discussed in Section~\ref{sec:vralgs}.  We are interested in the range of $\rho$ for which the VR algorithms will correctly reflect the true jet shape.  Note that the following results are only for guidance in choosing the value of $\rho$, and not a strict set of rules. 

In deriving the VR algorithms, we made use of the fact that distances in $\Delta S$ were linearly related to distances in $\Delta R$ via
\begin{equation}
\label{eq:vrmetric}
\Delta R \approx \Delta S \cosh \eta.
\end{equation}
Recalling the definition of pseudorapidity $\eta$ in terms of the polar angle $\theta$
\be
\cosh \eta = \frac{1}{\sin \theta}, \qquad  \left| \frac{d \theta}{d \eta} \right| = \sin \theta, 
\ee
it is easy to see that locally,
\begin{equation}
d S^2 \equiv d \theta^2+\sin^2 \theta \, d\phi^2 = \sin^2 \theta \,  (d\eta^2+ d\phi^2) = \sin^2 \theta \, d R^2.
\end{equation}
which is equivalent to \Eq{eq:vrmetric}.

The approximation in \Eq{eq:vrmetric} is valid as long as 
\begin{equation}
\left| \frac{d \theta}{d \eta} \right| \gtrsim \frac{1}{2}\Delta\eta \left| \frac{d^2\theta}{d\eta^2} \right|,
\end{equation}
which conservatively implies (taking $\Delta \phi \rightarrow 0$)
\be
\Delta R \lesssim 2 \left|\frac{d \theta}{d \eta} \right| \big/ \left| \frac{d^2\theta}{d\eta^2} \right| =  \frac{2}{\cos \theta }.
\ee
For the VR variants we are considering with $\Delta R = \rho/p_T$, distortions in the jet shape in going from $\Delta S$ to $\Delta R$ will be sufficiently small when
\be
\label{eq:restrictrho}
\rho \lesssim \frac{2 p_T}{\cos \theta}.
\ee

As a general rule of thumb then, we expect the VR algorithms to correctly reproduce the size of jets as long as $\rho\lesssim 2 p_T$  for most events reconstructed.  Because one must use a cutoff on $R_{\rm eff}$ to remove spuriously large cones, events which violate this condition are not necessarily grossly incorrect; they will be reconstructed as if by the CA or AKT algorithms working with $R_{\rm max}$.

\bibliography{jetbib}
\bibliographystyle{jhep}
\end{document}